\documentclass{aa}  

\usepackage{lscape}
\usepackage{graphicx}
\usepackage{array,multirow}
\usepackage{multirow}
\usepackage{txfonts}
\usepackage{gensymb}
\usepackage{natbib}
\bibpunct{(}{)}{;}{a}{}{,} 
\begin{document} 

\title{The Hydra\,I cluster core - II. Kinematic complexity in a rising velocity
dispersion profile around the cD galaxy NGC\,3311}

\author{M. Hilker\inst{\ref{eso}}
\and T. Richtler\inst{\ref{concepcion}}
\and C. E. Barbosa\inst{\ref{eso},\ref{usp}}
\and M. Arnaboldi\inst{\ref{eso}}
\and L. Coccato\inst{\ref{eso}}
\and C. Mendes de Oliveira\inst{\ref{usp}}}

\institute{European Southern Observatory, Karl-Schwarzschild-Stra\ss{}e 2,
 85748 Garching, Germany\label{eso}
\and
Universidad de Conc\'epcion, Conc\'epcion, Chile\label{concepcion}
\and
Universidade de S\~{a}o Paulo, Instituto de Astronomia, Geof\'isica e
 Ci\^encias Atmosf\'ericas, Rua do Mat\~ao 1226, S\~ao Paulo, SP,
 Brazil\label{usp}}

\date{Received August 8, 2017; accepted August 14, 2018}

 
\abstract
{NGC\,3311, the central galaxy of the Hydra\,I cluster, shows signatures 
of recent infall of satellite galaxies from the cluster environment. Previous
work has shown that
the line-of-sight velocity dispersion of the stars and globular clusters in the
extended halo of NGC\,3311 rises up to the value of the cluster velocity
dispersion. In the context of Jeans models, a massive dark halo with a
large core is needed to explain this finding. However, position dependent
long-slit measurements show that the kinematics are still not understood.} 
{We aim to find kinematic signatures of sub-structures in the
extended halo of NGC\,3311.}
{We performed multi-object spectroscopic observations of the diffuse
stellar halo of NGC\,3311 using VLT/FORS2 in MXU mode to mimic
a coarse `IFU'. The slits of the outermost masks reach out to about
35\,kpc of galactocentric distance. We use \textsc{pPXF} to extract
the kinematic information of velocities, velocity dispersions and the
high-order moments $h_3$ and $h_4$.}
{We find a homogeneous velocity field and velocity dispersion field
within a radius of about 10\,kpc. Beyond this radius, both the velocities and
the velocity dispersion start to depend on azimuth angle and show a
significant intrinsic scatter. The inner
spheroid of NGC\,3311 can be described as a slow rotator. Outside
10\,kpc the cumulative angular momentum is rising, however, without
showing an ordered rotation signal. If the radial dependence alone is
considered, the velocity dispersion does not simply rise but fills an
increasingly large range of dispersion values with two well defined
envelopes. The lower envelope is about constant at 200\,km\,s$^{-1}$.
The upper envelope rises smoothly, joining the velocity dispersion of the
outer globular clusters and the cluster galaxies. We interpret this behaviour 
as the superposition of tracer populations with increasingly shallower radial 
distributions between the extremes of the inner stellar populations and the
cluster galaxies. Simple Jeans models illustrate that a range of of mass
profiles can account for all observed velocity dispersions, including radial
MOND models.}
{The rising velocity dispersion of NGC\,3311 apparently is a result of
averaging over a range of velocity dispersions related to different tracer
populations in the sense of different density profiles and anisotropies.
Jeans models using one
tracer population with a unique density profile are not able to explain the 
large range of the observed kinematics. Previous claims about the cored
dark halo of NGC\,3311 are therefore probably not valid. This may in
general apply to central cluster galaxies with rising velocity dispersion
profiles, where infall processes are important.}

\keywords{galaxies: clusters: individual: Hydra I -- galaxies: individual:
 NGC\,3311 -- galaxies: halos -- Galaxies: elliptical and lenticular, cD --
 galaxies: formation -- galaxies: kinematics and dynamics}

\titlerunning{Kinematic complexity around the cD galaxy NGC\,3311}
\authorrunning{Hilker et al.} 

\maketitle
%

\section{Introduction}

The structure of galaxies can be strongly altered by environmental processes,
like major and minor mergers, accretion of satellite galaxies, ram pressure
stripping and other interactions between galaxies and with the intra-cluster
medium. 
A striking phenomenon in this respect is the existence of very extended
stellar halos around bright central galaxies in galaxy clusters \citep[e.g.][]
{1987ApJS...64..643S, 1988ApJ...328..475S}. Early classification efforts
assigned the term `cD' to those galaxies \citep{1964ApJ...140...35M,
1965ApJ...142.1364M} whose halos can in extreme cases embrace the entire
host galaxy cluster as in the case of Abell 1413 \citep{1976ApJ...209..693O,
2012A&A...548A..18C} frequently also presenting double or multiple nuclei
of central galaxies \citep[see][for a discussion of the early
literature]{1989ARA&A..27..235K}. In the following, we use the term `cD'
in this simple sense, irrespective of whether the halo can photometrically
identified as a separate entity or not, see \citet{2015ApJ...807...56B} for
a more profound discussion.

The cluster environment is clearly important for the build-up of such a
large halo, either by coalescence of larger galaxies or by the infall of
smaller galaxies or tidal debris of cluster material. This is plausibly also
the reason for the immense richness of the globular cluster systems (GCSs)
of central galaxies \citep[e.g.,][]{2017ApJ...835..101H} and, in the case of
this study, the rich GCS of NGC\,3311, the central galaxy of the Hydra\,I
galaxy cluster \citep{2008ApJ...681.1233W}.

Given that the halos of massive ellipticals in general grow by a factor of
about 4 in mass since $z=2$ \citep{2010ApJ...709.1018V}, one expects
an even higher growth rate in the centres of galaxy clusters, where the
galaxy number density is highest.
The more recent mass growth is dominated by the accretion of low mass
systems (minor mergers) which may leave kinematical signatures in the
phase space of the outer stellar population of central cluster galaxies.
These extreme environments, therefore, are suitable to study the main
physical processes that cause the destruction of infalling galaxies as well as
the build-up of the intra cluster light and the central dark matter halo
of a galaxy cluster.

The extended envelopes make it possible to probe the kinematical properties
of the galaxy light out to large radii. A rise of the velocity dispersion of the
galaxy light is not unusual for the most massive early-type galaxies 
\citep{2013ApJ...765...24N, 2017MNRAS.464..356V}. The first measured
rising velocity dispersion profile was reported by \citet{1979ApJ...231..659D}
for the cD galaxy IC\,1101 in the cluster Abell\,2029. In the second known
and well studied case of NGC\,6166, the velocity dispersion stays constant
at a value of 200\,km/s until a radius of 10\,kpc and then steeply rises to
high values above 800\,km/s at large radii, reaching the velocity
dispersion of galaxies in the cluster \citep{1999MNRAS.307..131C,
2002ApJ...576..720K, 2015ApJ...807...56B}.

One of the nearest cD-galaxies is NGC\,3311, the central galaxy in the
Hydra\,I galaxy cluster. Recent publications on NGC\,3311 are
mainly from our group \citep{2008AN....329.1057V, 2010A&A...520L...9V,
2011A&A...528A..24V, 2011A&A...533A.138C, 2011A&A...531A...4M,
2011A&A...531A.119R, 2012A&A...545A..37A, 2016A&A...589A.139B,
2018A&A...609A..78B}.
Photometric and kinematical studies using the galaxy light, planetary nebulae
(PNe) and globular clusters (GCs) clearly demonstrate the close connection of
the inner galaxy with its host cluster. Asymmetric light distribution and tidal
tails are evidences of recent infall processes \citep{2012A&A...545A..37A}.
Planetary nebulae show a non-Gaussian velocity distribution with peaks that
are offset from the systemic velocity \citep{2011A&A...528A..24V}. The central
velocity dispersion of the galaxy light is only 150\,km/s (but fits to its low
surface brightness) and rises up to almost 400\,km/s within 10\,kpc. The more
distant globular clusters reach an even higher velocity dispersion, equal to
that of the cluster galaxies \citep{2011A&A...531A...4M, 2011A&A...531A.119R} .
To explain this rise with a simple Jeans model, one needs a large core of
dark matter.
The population composition of NGC\,3311 indicates a distinction
between the original elliptical galaxy and the later accreted material: the
inner galaxy is old and metal-rich, while at larger radii the scatter in
metallicity becomes considerably larger \citep{2016A&A...589A.139B}.

Apparent discrepancies of kinematical values, in particular the velocity
dispersion, measured with long-slits of different
azimuthal orientations have been reported by \citet{2011A&A...531A.119R}.
In this paper we demonstrate that indeed the velocity and velocity dispersion
show a two-dimensional irregular pattern and that the kinematics at
radii larger than 10\,kpc cannot be described by a radial coordinate only.
This continues the trend already seen in the high surface brightness parts
of the galaxy, where the dynamical analysis of MUSE observation has
revealed an asymmetric velocity and velocity dispersion field with azimuthal
variations  \citep{2018A&A...609A..78B}. The present work emphasizes that
it is important to investigate the entire velocity field of the galaxy light and its
kinematic properties out to large radii. The database of the present paper
has already been used in \citet[][Paper\,I]{2016A&A...589A.139B} to analyse
the stellar population properties of NGC\,3311.

The present paper is structured as follows.
In Section \ref{sec:obsred} we present the detailed observational properties and
data reduction. The kinematical analysis is described in Section \ref{sec:obskin}.
The measurement of stellar kinematics and main results are presented and
discussed in Sections \ref{sec:results} and \ref{sec:discussion}, with the final
summary and conclusion in Section \ref{sec:conclusion}. Throughout this paper,
we adopt the distance to the Hydra\,I cluster core as 50.7\,Mpc. 


\section{Observations and data reduction}
\label{sec:obsred}

\subsection{Observations}
\label{sec:obs}

For our study of NGC\,3311's halo we used data from ESO programme 088.B-0448
(PI: Richtler). The spectroscopic science data were taken in seven different nights 
between January 17th and March 29th, 2012, with FORS2 (in multi-object mode,
using the Mask eXchange Unit (MXU)) mounted on UT1 at Paranal Observatory.
We used the 1400V grism, which
covers the wavelength range 4600-5800\AA. With a slit width of 1\arcsec and a 
dispersion of 0.31\AA/pixel we reach a spectral resolution of $R=$2100 at
5200\AA, which translates into a velocity resolution of $\sim$140\,km/s.
Six different MXU masks were necessary to cover the halo around NGC\,3311.
For each mask twilight skyflats were taken, mostly in the same nights as the
science exposures. Those flats are important to calibrate the relative responses
between object and sky slits.

The mask design follows an 'onion-shell' approach. Each mask consists of a
half-ring of halo slits, all positioned at about the same galactocentric distance
from  NGC\,3311.
The typical dimensions of the slits are 1$\times$5\arcsec. The combination of 
2$\times$3 masks with half-rings of slits at different galactocentric distances on
each side of NGC\,3311 complete the full onion-shell of halo slits. This is shown in
Figure\,\ref{fig:findchart}. The outermost halo slits reach distances of 150\arcsec 
($\sim35$\,kpc) from the galaxy centre. The onion-shell approach allows us to
study the stellar population properties at several radii as well as at different
azimuthal angles in the halo of NGC\,3311. It kind of mimics a coarse IFU.
The advantage of such an approach is that an `effective IFU field-of-view'
of about $4\times4$ arcmin is homogeneously covered, which is $16\times$
larger than the so far largest single IFU Multi-Unit Spectroscopic Explorer
(MUSE) at the VLT.

The halo slits avoid point sources and galaxies, except the lenticular galaxy HCC\,007
south of NGC\,3311. The position angle of the masks and thus slits was chosen such
that we have free sky regions close to the borders of the 7.8\arcmin field-of-view in
spatial direction of the slits. This is the case north-east and south-west of
NGC\,3311, as can be seen in Figure\,\ref{fig:findchart}. We distributed the sky slits
such that they are located at the same x-positions (in CCD coordinates) as the halo
slits. This guarantees that we always have a sky spectrum that covers the same
wavelength range as the corresponding halo spectrum. We named the slits in the
different half-shells with the prefixes 'cen1', 'cen2', 'inn1', 'inn2', 'out1' and 'out2'
from inwards out, and numbered them with increasing spatial position on the y-axis
of the CCD chip (see Figure\,\ref{fig:findchart}).

\begin{figure}[t]
\centering
\includegraphics[width=0.95\linewidth]{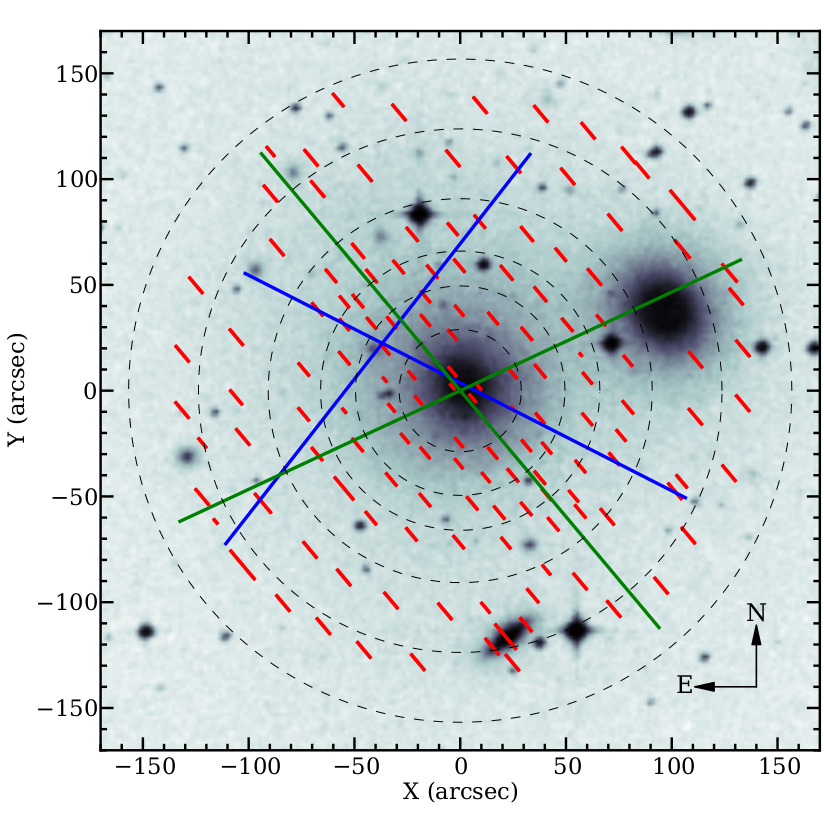}
\caption{Positioning of the slits in the halo of NGC\,3311. The small red slits are
our novel set of observations, while the two blue and green long slits represent
the positioning of previous work by \citet{2010A&A...520L...9V} and
\citet{2011A&A...531A.119R}, respectively. One of the green long slits crosses
the neighbouring giant elliptical NGC\,3309 north-west of NGC\,3311.}
\label{fig:slits}
\end{figure}

The exposure times of the two innermost shells (cen1 and cen2 and inn1 and
inn2) were 2$\times$1400 seconds. The masks of the outer shells (out1 and
out2) were exposed  6$\times$1400 seconds. Between the individual exposures
a dither along the slit of up to $\pm0.6$\arcsec was applied.

Figure\,\ref{fig:slits} shows our onion-shell halo slits (in red) in the context of
previous long-slit observations. The blue and green slits belong to data presented
in \citet{2010A&A...520L...9V} and \citet{2011A&A...531A.119R}, respectively. 
Our slit design resulted in 135 slits dedicated to the halo of NGC\,3311 and
NGC\,3309. Not all of them could be used because the 1400V grism causes a
tilt of the light beam which shifts the spectrum on the detector in Y direction
by $\sim$28\arcsec. In this way three spectra, one in each of the masks 'cen1',
'cen2' and 'inn1', were lost because they ended up in the chip gap of the
two FORS2 CCDs. And in the other masks six slits, two in each mask, were 
shorter than expected because of the gap, and thus the spectra of those slits
have a slightly lower S/N than other spectra at similar surface brightnesses.

\subsection{Data reduction}
\label{sec:red}

The reduction was carried out with custom scripts based on \textsc{IRAF} 
tasks, mainly taken from the \textsc{twodspec}, \textsc{longslit},
\textsc{apextract} and \textsc{onedspec} packages \citep[e.g.][]{1993ASPC...52..173T}. 
All science, twilight skyflat and arc lamp exposures were first bias and flatfield
corrected. Individual twilight skyflats of the same mask were combined. The
science exposures were cleaned from cosmics using the \textsc{lacos\_spec}
routine from \citet{2001PASP..113.1420V}. We determined the spatial
distortion of the spectra by identifying and tracing the slit gaps in the
normalized flatfield as `absorption features' (using the \textsc{IRAF} tasks
\textsc{identify}, \textsc{reidentify} and \textsc{fitcoords}). The science
exposures, twilight flats and arc lamp exposures were then straightened
by applying the derived coordinate transformation. The spectral regions
of all individual slits were cut out in all rectified frames. An illumination
correction of the science exposures and twilight skyflats along the slits was
applied, which was derived from the twilight skyflats.
Next, a 2-dimensional wavelength calibration was performed on
the arc lamp exposures of each slit and applied to the science exposures.
The position of the 5577.3\AA [O\,{\sc I}] sky line was used to
correct for residual zeropoint shifts in the wavelength calibrations, which
were of the order of 0.1-0.4\AA. The individual calibrated science exposures
of each mask were then averaged.
We interactively used the \textsc{IRAF} task \textsc{apall} to define the
aperture to be extracted in the individual slits, avoiding edge effects. The
same extraction was applied to the twilight skyflats.
The very central slits of NGC\,3311 as well as the three slits in HCC\,007
have a very high S/N, and we have split them into 2 (central) and 3 and 5
(HCC\,007 outer and inner) sub-apertures, respectively.

\begin{figure}[t]
\centering
\includegraphics[width=0.98\linewidth]{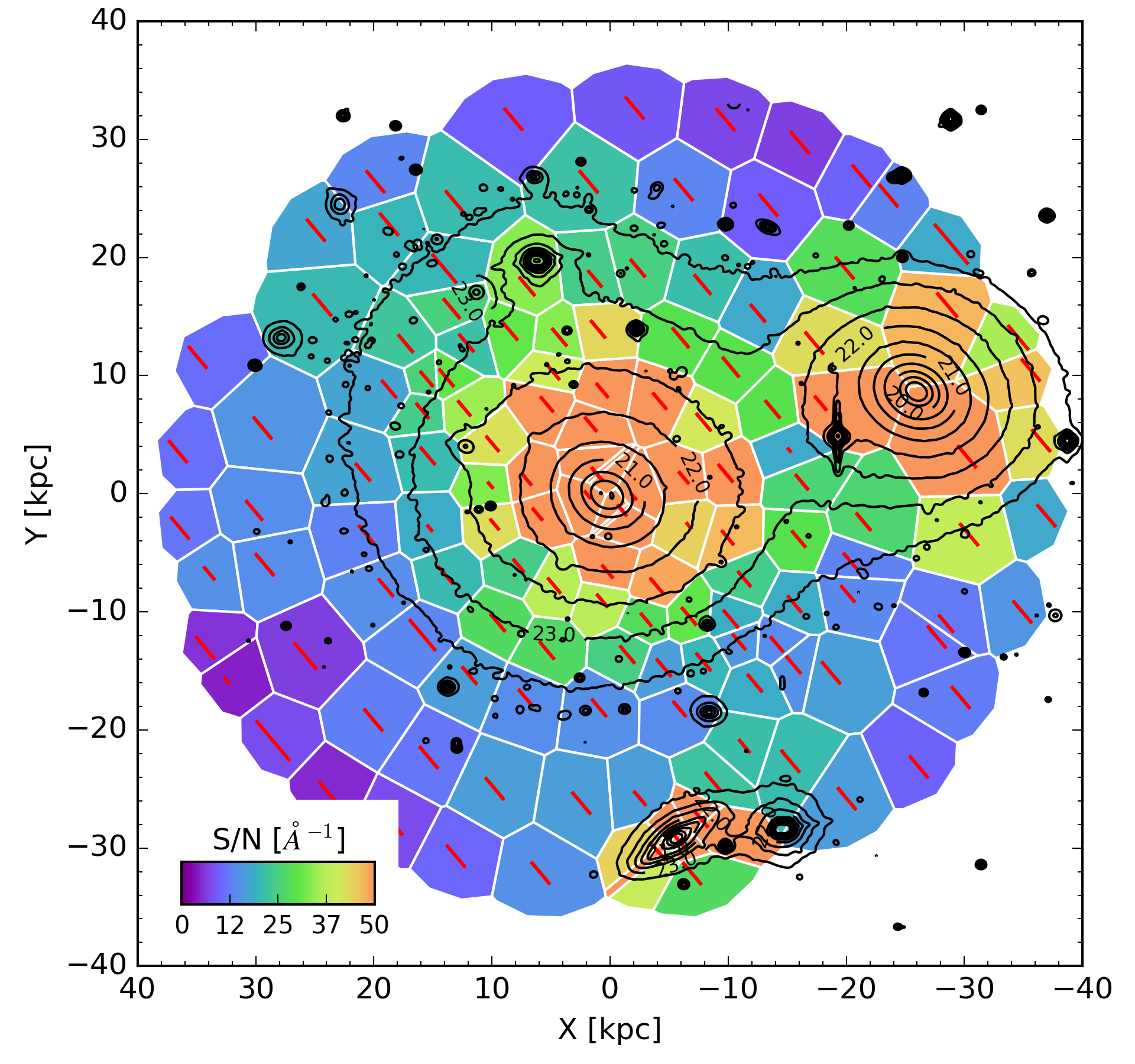}
\caption{Map of the signal to noise in each region after Voronoi
Tesselation of the whole slit area. The contours show the $V$-band
surface brightness levels of objects in the field-of-view ranging
from $\mu_V=20$ to 23.5 mag\,arcsec$^{-2}$ in steps of 0.5
mag\,arcsec$^{-2}$, from \citet{2012A&A...545A..37A}.} 
\label{fig:signaltonoise}
\end{figure}

\begin{figure*}[t]
\centering
\includegraphics[width=0.331\linewidth]{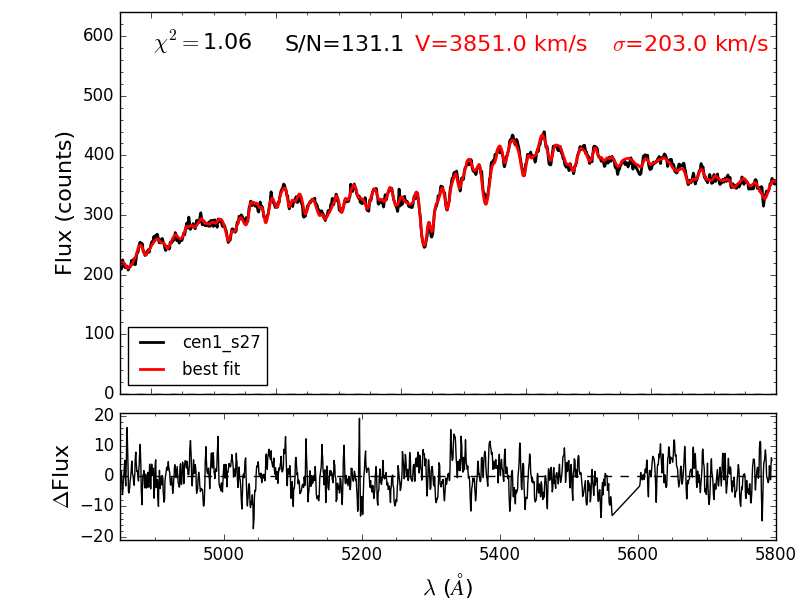}
\includegraphics[width=0.331\linewidth]{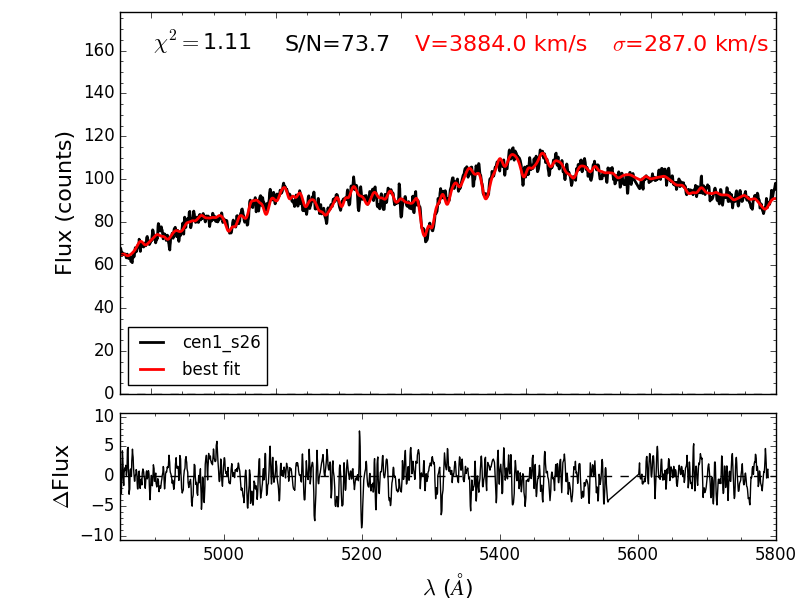}\includegraphics[width=0.331\linewidth]{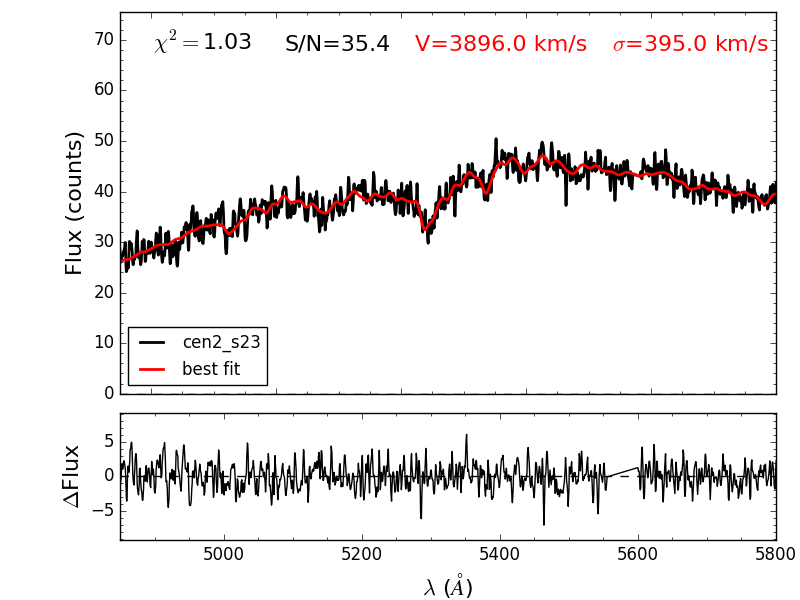}
\includegraphics[width=0.331\linewidth]{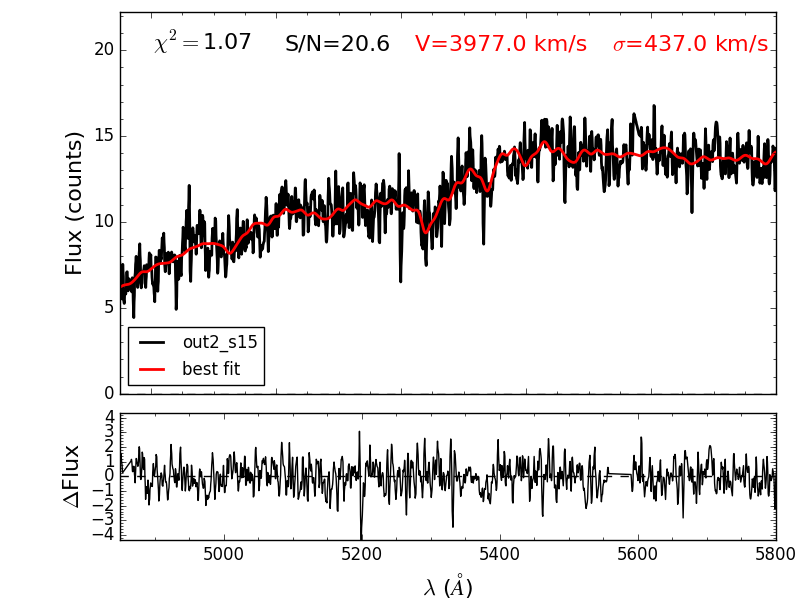}
\includegraphics[width=0.331\linewidth]{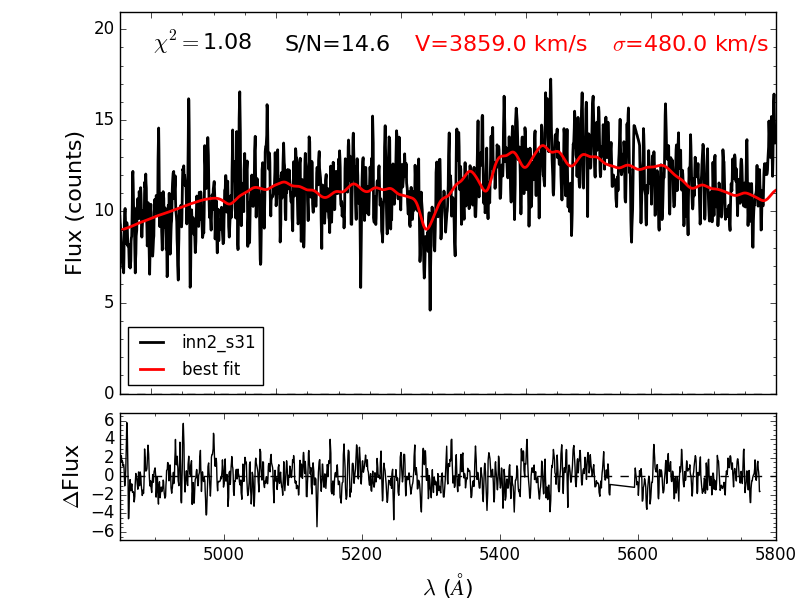}\includegraphics[width=0.331\linewidth]{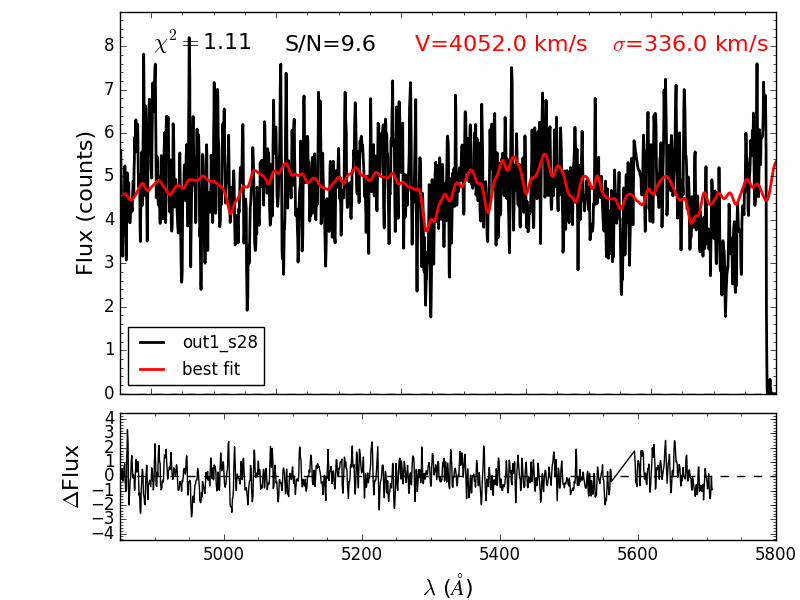}
\caption{Here we show six example spectra from high to low S/N (from
top left to bottom right) together with their \textsc{pPXF} fits (red lines in
upper panels) and residuals (lower panels). Apart from the S/N value, the
name of the spectrum, the derived radial velocity and velocity dispersion
as well as the reduced $\chi^2$ value of the fit are given as legend in
each panel.}
\label{fig:example_spectra}
\end{figure*}

The last, most important step was to subtract the sky from the galaxy
spectra, taking into account the different sizes and sensitivities of the
galaxy-sky slit pairs. For that we collapsed for each slit the 2-dimensional
twilight
skyflats of the galaxy and sky slits by averaging all lines along the spatial
direction. The 1-dimensional twilight spectra of galaxy and sky were
divided by each other to derive the relative response between galaxy and
sky slit. We fitted a spline of 7th order to the response spectrum, and
used this fit to bring the sky spectrum to the same sensitivity as the
galaxy spectrum. Only then, we subtracted the sensitivity corrected sky
spectrum from the galaxy spectrum. Remaining sky residuals of the 
5577.3\AA [O\,{\sc I}] sky line were masked out in the final 
1-dimensional galaxy spectra.
Finally, all spectra were corrected for having their wavelength scale in
heliocentric velocities, that is the motion of the Sun with respect to the 
observed direction as calculated from the header parameters was taken
out.

We did not perform flux calibration because this is not straightforward
for FORS2/MXU spectra. The standard star is taken in the middle of the
CCD, whereas the MXU slits are distributed all over the CCD. Since the
response function depends on the x-position on the CCD and all slits
cover different wavelength ranges, most of the extracted spectra would
get a slightly wrong and incomplete flux calibrations. In any case, for the
purpose of this paper, that is to derive the kinematics of NGC\,3311's halo, 
a flux calibration is not needed.

\begin{figure*}
\centering
\includegraphics[width=1.00\linewidth]{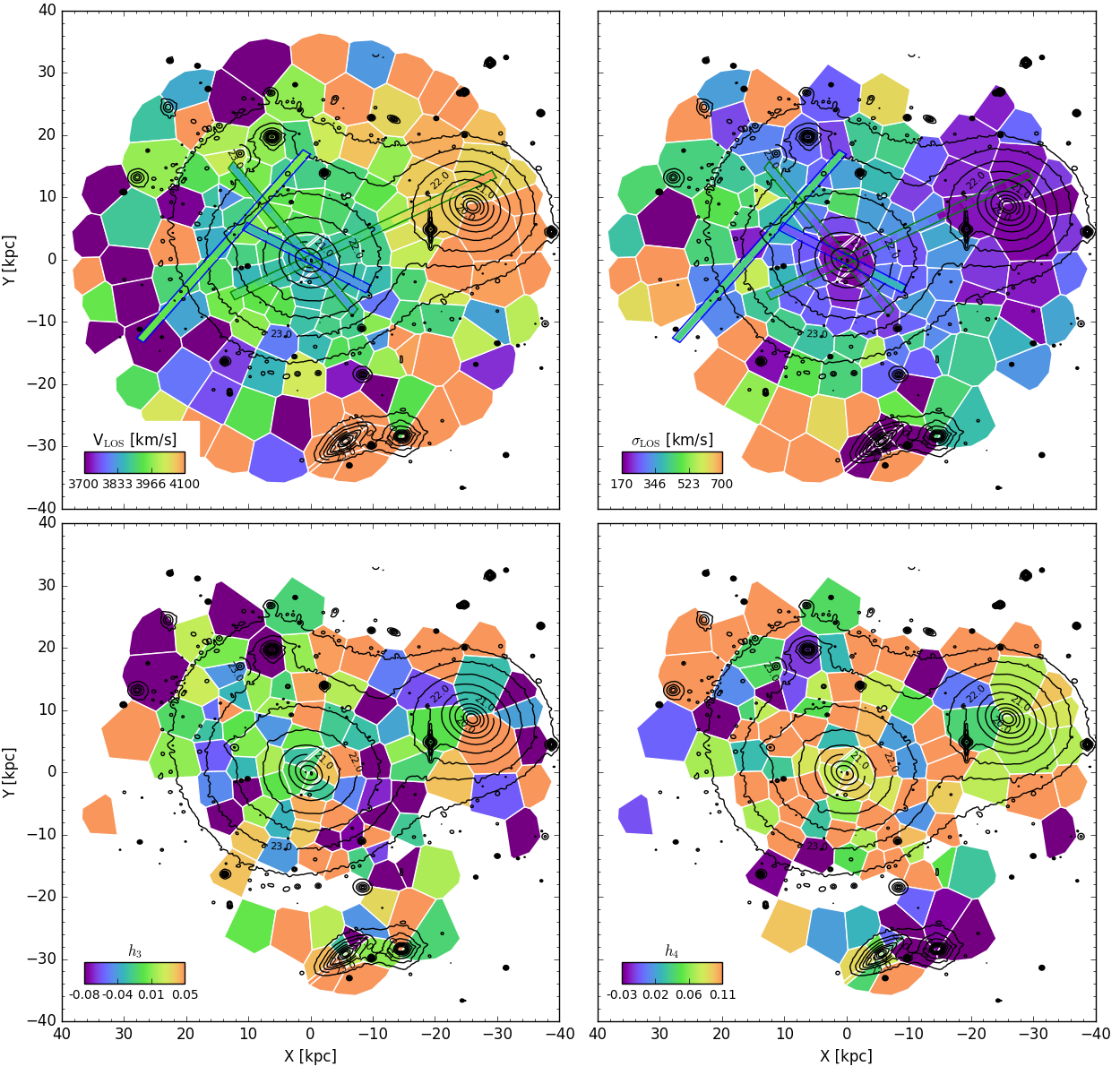}
\caption{Kinematic maps of the core of the Hydra\,I cluster. Upper left:
Line-of-sight recession velocity, $V_{\rm LOS}$ for spectra with S/N$>$5
per \AA. The results of previous long-slit data \citep{2010A&A...520L...9V,
2011A&A...531A.119R} are shown as well. Upper left: Line-of-sight velocity
dispersion, $\sigma_{\rm LOS}$. Only results with S/N$>$10 per \AA are
shown. Bottom left: Higher moment $h_3$ (skewness) for data with S/N$>$15.
Bottom right: Higher moment $h_4$ (kurtosis) for data with S/N$>$15.
In all maps, $V$-band contours are shown in black, ranging from 20 to 23.5
mag\,arcsec$^{-2}$ in steps of 0.5 mag\,arcsec$^{-2}$, from
\citet{2012A&A...545A..37A}. In all panels, North is up and East is left.}
\label{fig:kinmaps}
\end{figure*}

\section{Stellar kinematics}
\label{sec:obskin}

We measured the line-of-sight (LOS) stellar kinematics of our dataset using
the penalized pixel-fitting code \citep[\textsc{pPXF,}][]{2004PASP..116..138C},
which allows the simultaneous fitting of a linear combination of stellar
templates with a line-of-sight velocity distribution (LOSVD). We carefully selected
the fitting ranges for each slit individually by eye to avoid continuum
discontinuities, sky lines and strong noise. The overall maximum wavelength
range for all spectra is 4510\AA$\lesssim\lambda\lesssim$5850\AA, but
the selected fitting regions vary depending on the slit positions
on the chip, and thus the wavelength region covered. The common minimum
range of fitting regions for all spectra is 4900\AA$\lesssim\lambda\lesssim$5520\AA.
For consistency, we calculated the S/N per \AA for all spectra in the 
wavelength range 5200\AA$\leq\lambda\leq$5500\AA. We measured the
S/N by comparing the integrated light in the observed spectrum with the
standard deviation of the residuals between the observed spectrum and the
best fit template from \textsc{pPXF}.  A map of the S/N values per spectrum
is shown in Figure\,\ref{fig:signaltonoise}.

\begin{figure*}
\centering
\includegraphics[width=1.00\linewidth]{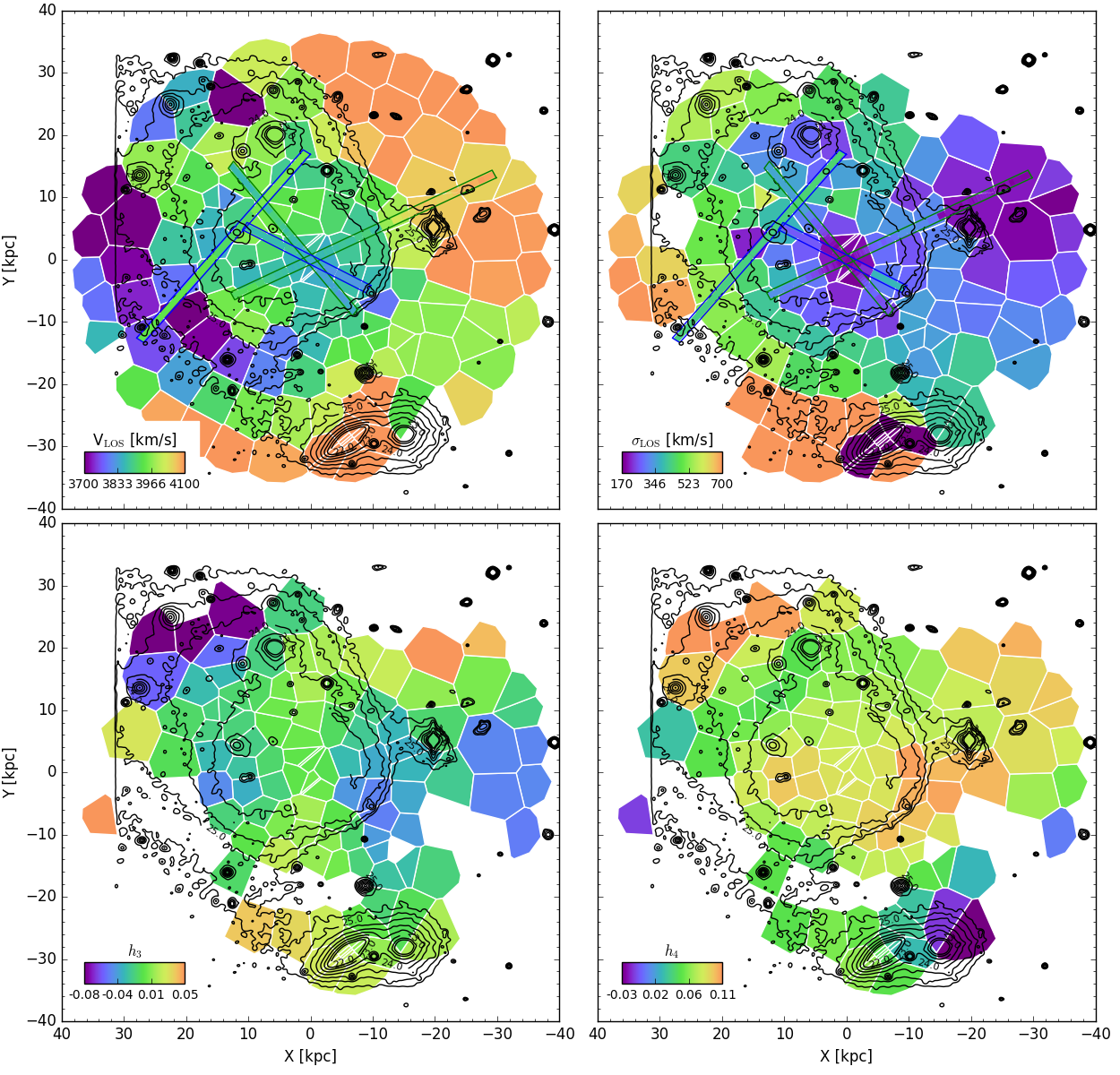}
\caption{Same as in Figure\,\ref{fig:kinmaps}, except that data are
smoothed (via the LOESS algorithm, see text) for regions of S/N$<$20
in the upper two panels and for all regions in the lower two panels.
Moreover, the black surface brightness contours here show the
$V$-band residual map, ranging from 22 to 26 mag\,arcsec$^{-2}$
in steps of 0.5 mag\,arcsec$^{-2}$, after the subtraction of maximum
symmetric models from the two central galaxies NGC\,3311 and
NGC\,3309 \citep{2012A&A...545A..37A}. In all panels, North is up
and East is left.}
\label{fig:kinmaps_loess}
\end{figure*}

As templates for the spectral fitting we took single stellar population spectra
from \citet{2010MNRAS.404.1639V}, which uses spectra from the MILES
stellar library \citep{2006MNRAS.371..703S}, computed with a resolution
of 2.5\AA, a Salpeter initial mass function (IMF) with logarithmic slope of 
$-1.3$, ages from 0.1 to 15 Gyr and metallicities in the range $-2.3 \leq$
[Z/H] $\leq 0.2$ dex. We convolved our observed spectra with a Gaussian
filter to bring the instrumental resolution of FWHM$=$2.1\AA to the resolution
of the MILES template spectra of FWHM$=$2.5\AA.

For the fitting, we used a velocity distribution characterised by the
velocity ($V_{\rm LOS}$), the velocity dispersion ($\sigma_{\rm LOS}$)
and the two high order Gauss-Hermite moments $h_3$ and $h_4$, which
measure deviations from the simple Gaussian profile. $h_3$ describes the
skewness of the Gaussian, in other words negative values indicate a distribution
with an extended tail towards low velocities, and positive values vice versa.
$h4$ describes the kurtosis of the Gaussian, that is positive values describe a
distribution more 'peaked' than a Gaussian, and negative values indicate a
more flat-topped distribution. There is no simple relation between $h_4$
and the anisotropy. Apart from the anisotropy, $h_4$ depends on the
potential and the density profile of the tracer population. An exact Gaussian
is produced by an isotropic population that generates an isothermal potential.

Besides the best fitting parameters, \textsc{pPXF} also provides a noiseless
best fit spectrum, and the residual difference between the observed 
and best fit spectrum gives a good estimation of the observation noise.
Uncertainties for the four parameters were calculated via Monte Carlo
simulations based on the S/N for each spectrum. 
However, \textsc{pPXF} makes a global minimization in the parameter
space with a single figure-of-merit function ($\chi^2$), and in some 
cases the `best' parameters are not a good description of the overall
shape of the spectra. This issue, the so-called template mismatch, can
be avoided in most cases by careful visual inspection of the fitting results.
We, therefore, thoroughly tested, in an interactive manner, which
combination of additive and multiplicative polynomials for the
continuum fit in \textsc{pPXF} ($adegree$ and $mdegree$,
respectively) gave the most reliable results. We chose $adegree=6$
and $mdegree=4$ for most of the spectra, and $adegree=8$ and
$mdegree=2$ for some exceptions. The results of kinematic
measurements with \textsc{pPXF} are presented in 
Table\,\ref{tab:kinematics} in the appendix, and six example spectra
with different S/N values are shown in Figure \ref{fig:example_spectra}
together with their best \textsc{pPXF} fits and residuals.

We produced 2D maps of the kinematic parameters extrapolating the slits
using a Voronoi tesselation (see Figure\,\ref{fig:kinmaps}). In all maps the
same polygons as shown in the S/N map (Fig.\,\ref{fig:signaltonoise}) are
used. We applied different minimum S/N cuts to the four parameters, which
are S/N$>$5 for $V_{\rm LOS}$, S/N$>$10 for $\sigma_{\rm LOS}$, and
S/N$>$15 for the higher moments $h_3$ and $h_4$. Those cuts were
motivated by visual inspection of the fitting results. Below these cuts the
derived values were considered unreliable.

In order to improve the visualization of the main trends in the maps
and to provide a kind of binned 2D map with higher S/N, we
also produced filtered maps obtained with a locally weighted scatterplot
smoothing (LOESS) algorithm \citep{1979cleveland} calculated with the
Python implementation of \citet{2013MNRAS.432.1862C}. This non-parametric
method is suitable for the reconstruction of functions on noise data, and 
has a single free parameter, the fraction of data points to be considered
around each data point. We set this fraction to 0.3 for all our maps.
For the $V_{\rm LOS}$ and $\sigma_{\rm LOS}$ maps only regions with
S/N<20 were smoothed, whereas higher S/N polygons show their original
values because their $V_{\rm LOS}$ and $\sigma_{\rm LOS}$ values are
highly reliable due to small relative errors. For the $h_3$ and $h_4$ maps all
polygons were smoothed, thus increasing the signal of structures in adjacent
polygon cells of the 2D maps and overcoming the large relative errors of h3
and h4. The smoothed maps illustrate that the small scale variations in the
kinematic values do not only depend on the scatter of low S/N bins.

Finally, for the analysis of our results, we overplotted on these maps two
sets of surface brightness contours. The first set was produced from a
$V$-band image in surface brightness intervals ranging from 20 to 23.5
mag\,arcsec$^{-2}$ in steps of 0.5 mag\,arcsec$^{-2}$ (see
Figure\,\ref{fig:kinmaps}). The second set was produced from the residual
$V$-band image after maximum symmetric models were subtracted from
the two central galaxies NGC\,3311 and NGC\,3309
\citep{2012A&A...545A..37A}. The contours range from 22 to 26
mag\,arcsec$^{-2}$ in steps of 0.5 mag\,arcsec$^{-2}$ (see
Figure\,\ref{fig:kinmaps_loess}).
This residual image clearly shows the non-symmetrical excess of light
in the north-east region of NGC\,3311, as well as some excess of light
close to the lenticular galaxy HCC\,007 at the bottom of the image. As
one can see in Figure\,\ref{fig:signaltonoise}, also the S/N of our spectra
is enhanced in these excess regions.

\section{Results}
\label{sec:results}

\begin{figure*}
\centering
\includegraphics[width=0.415\linewidth]{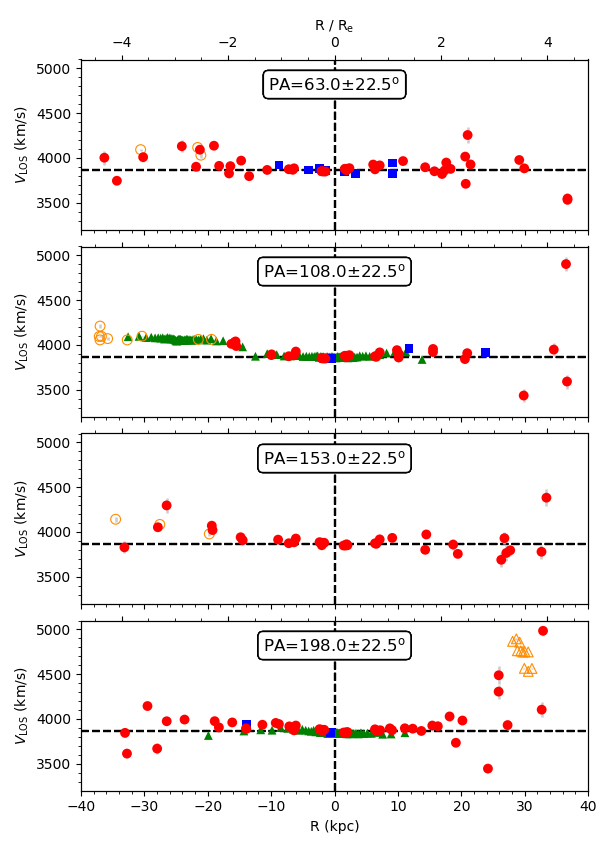}
\includegraphics[width=0.165\linewidth]{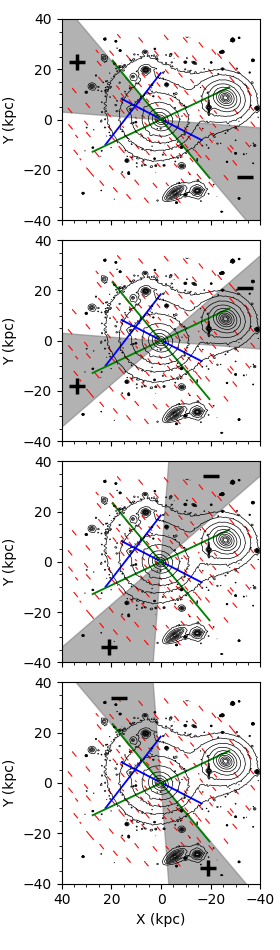}\includegraphics[width=0.415\linewidth]{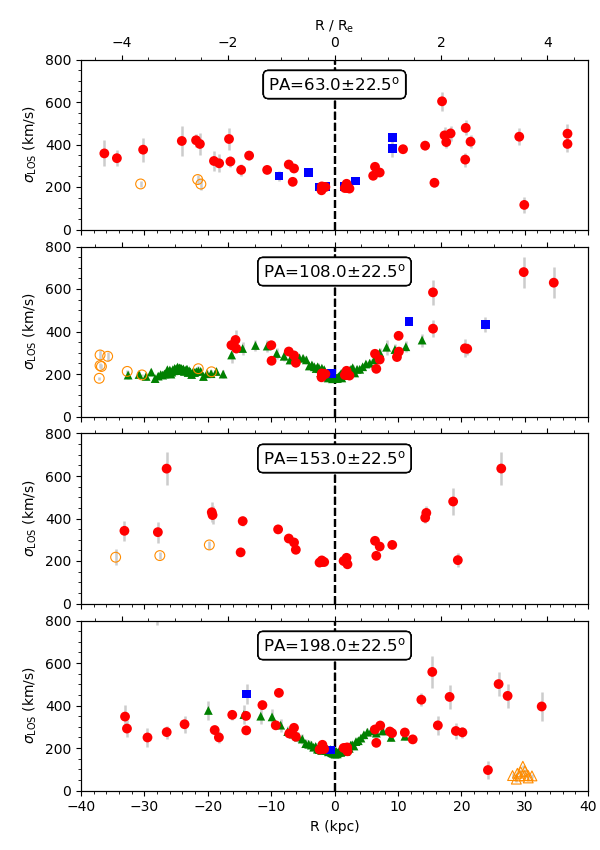}
\caption{Left: Line-of-sight velocity as function of galactocentric
distance in kpc from NGC\,3311 for four conic segments. Red dots
and orange open symbols are our measurements, blue squares
and green triangles those from long-slit data of \citet{2010A&A...520L...9V}
and \citet{2011A&A...531A.119R}. The orange circles are measurements
dominated by the light of NGC\,3309, open triangles those of HCC\,007.
The position angles, PA (North over East) are indicated as legend in the
panels. The horizontal dashed line indicates the systemic velocity of
NGC\,3311 of 3850\,km\,s$^{-1}$.
Middle: Grey areas show the cones, in which slits have been
selected for the left and right plots. Right: Same as left but for the
line-of-sight velocity dispersion. In the left and right panels the
distance in units of effective radii is given on the top label.
We note that a positive distance in the x-axis of these panels refers to the
direction of the position angles as indicated in the plots (North over
East), the negative x-axis refers to the opposite PA direction).
The axes description of the middle panels are distances in RA and
DEC (positive towards North and East, negative towards South and
West). The directions of the radial distances along the cones
shown in the left and right panels are indicated as black '+' and '-' signs.}
\label{fig:velsig_profiles}
\end{figure*}

\begin{figure*}
\centering
\includegraphics[width=0.415\linewidth]{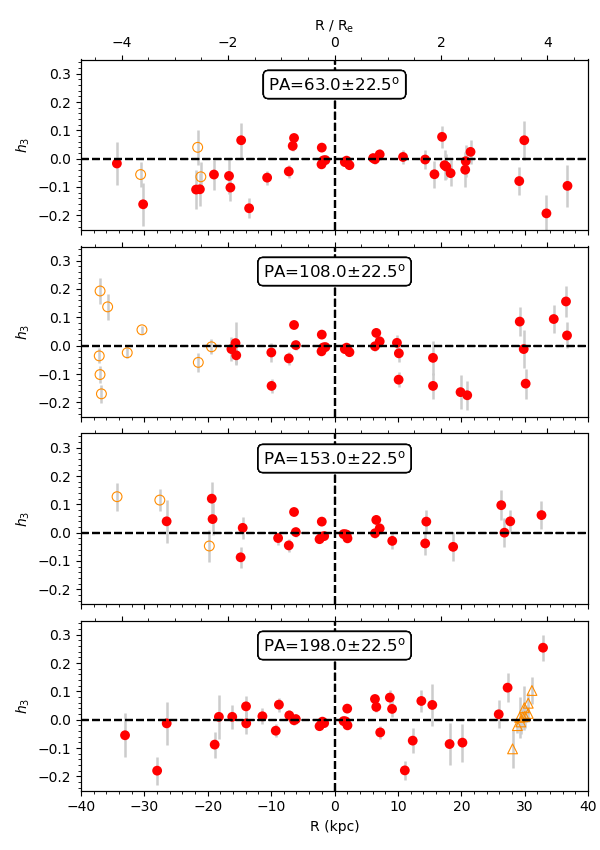}
\includegraphics[width=0.165\linewidth]{cones.png}\includegraphics[width=0.415\linewidth]{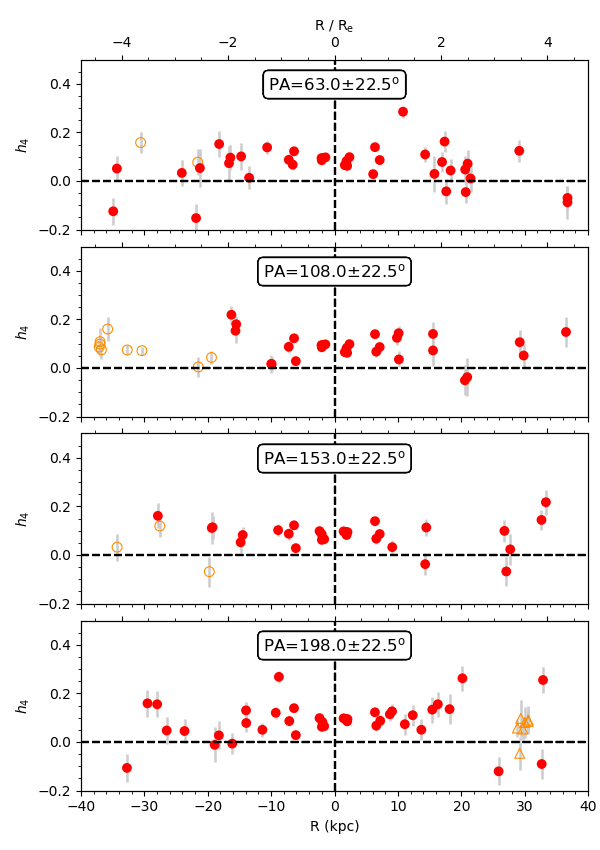}
\caption{Same as Figure \ref{fig:velsig_profiles} for the
Gauss-Hermite coefficients $h_3$ (left) and $h_4$ (right).}
\label{fig:h3h4_profiles}
\end{figure*}

The final kinematic maps are presented in Figures\,\ref{fig:kinmaps}
and \ref{fig:kinmaps_loess}. Colour bars in the legend indicate the 
parameter range displayed in the individual maps.
In order to analyse kinematic trends with galactocentric distance from
NGC\,3311 and position angle around it we created position-$V_{\rm LOS}$
and position-$\sigma_{\rm LOS}$ diagrams in four different segments
of 45 degree width, see Figure\,\ref{fig:velsig_profiles}. The position angle,
PA, is defined as North$=0^{\rm o}$ and East$=90^{\rm o}$. The four
PA values are chosen such that they coincide with the galaxy's major axis
\citep[$63^{\rm o}$,][]{2012A&A...545A..37A}, its minor axis ($153^{\rm o}$)
and the two orientations of the long-slits at $108^{\rm o}$ and
$198^{\rm o}$ from \citet{2011A&A...531A.119R}, see left panels in
Figure\,\ref{fig:velsig_profiles}. Radial bins are also given in units of
effective radii (top labels), with $R_e= 8.4$\,kpc for NGC\,3311
\citep{2012A&A...545A..37A}. We corrected for systematic velocity offsets
between the long-slit and our results, as discussed below.
We also produced similar plots for the Gauss-Hermite coefficients $h_3$
and $h_4$, see Figure\,\ref{fig:h3h4_profiles}. In all these plots we only show
data (Voronoi tesselation cells) that fulfil the following error selection for the
respective quantity ($V$, $\sigma$, $h_3$ and $h_4$):
$\sigma(V_{\rm LOS})<100$\,km\,s$^{-1}$,
$\sigma(\sigma_{\rm LOS})<80$\,km\,s$^{-1}$, $\sigma(h_3)<0.1$ and
$\sigma(h_4)<0.1$.

\begin{figure*}
\centering
\includegraphics[width=0.405\linewidth]{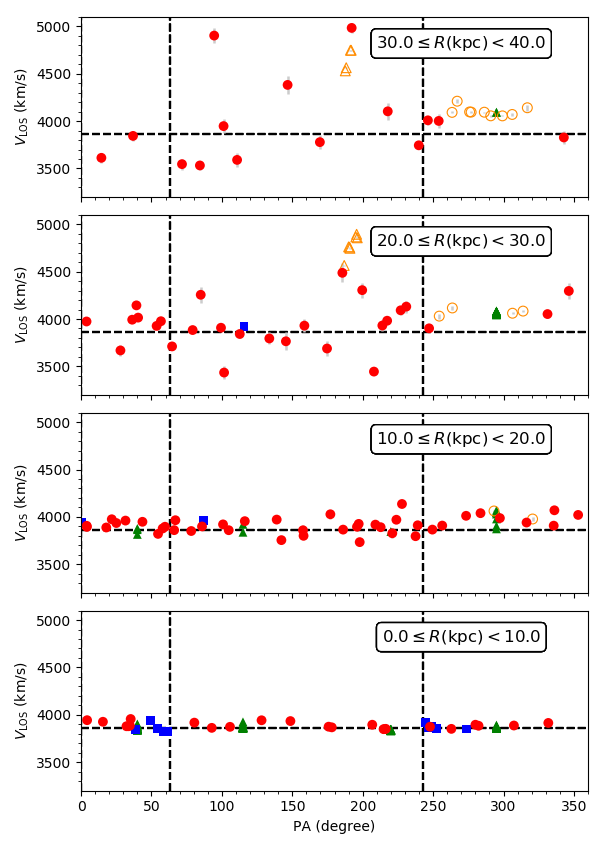}
\includegraphics[width=0.170\linewidth]{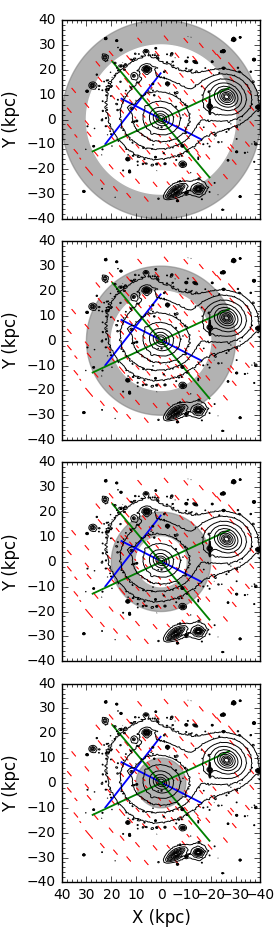}
\includegraphics[width=0.405\linewidth]{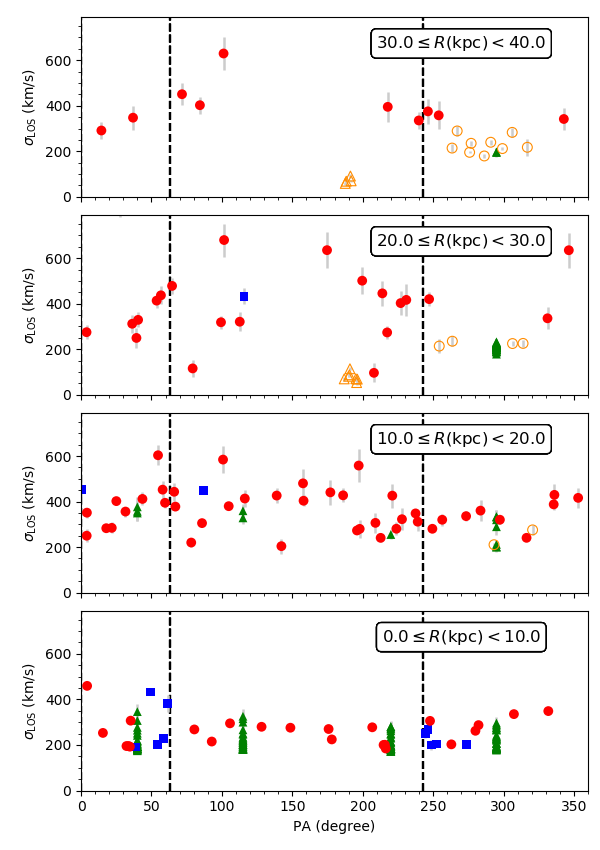}
\caption{Left: Line-of-sight velocity as function of the position angle,
PA (North over East), for four radial bins of 10\,kpc width between
zero and 40\,kpc galactocentric distance from NGC\,3311, as
indicated in the legend of the panel. The radial bins increase from
bottom to top. The symbols are the same as in
Figure\,\ref{fig:velsig_profiles}. Middle: Grey areas show the
annular areas, in which slits have been selected for the left and
right plots.Right: Same as left, but for the line-of-sight velocity
dispersion. In the left and right panels the vertical dashed lines
indicate the galaxy's major axis.}
\label{fig:velsig_pa}
\end{figure*}

Moreover, we present in Figure\,\ref{fig:velsig_pa} $V_{\rm LOS}$ and
$\sigma_{\rm LOS}$ as function of position angle around NGC\,3311
in four radial bins of 10\,kpc width from the centre out to 40\,kpc.
The distributions show the degree of symmetry and scatter of the
LOSVD moments as function of distance to the galaxy centre
out to $\sim$5 effective radii.
In Figures\,\ref{fig:velsig_profiles} to \,\ref{fig:velsig_pa}, those slits that
are dominated by light from the giant elliptical NGC\,3309 and the lenticular
galaxy HCC\,007 are indicated by open orange symbols (circles and triangles,
respectively). The selection of those slits was guided by the surface brightness
contours shown in Figure\,\ref{fig:signaltonoise}. In Table\,\ref{tab:kinematics} 
the spectra that belong to the halo of NGC\,3311, NGC\,3309 and HCC\,007
are identified by the flags 1, 2, and 3, respectively, in column (13).

Additionally, in order to highlight the asymmetric distribution of the velocity
and velocity dispersion, we plot in Figures\,\ref{fig:velocity_folded} and 
\ref{fig:dispersion_folded} folded values for four position angles,
overplotting the results for the positive major-axis on top of the
results from the negative major-axis. We exclude from this plot the
data points around NGC\,3309 and HCC\,007, both for our data and the
longslit data, to only show the profiles of NGC\,3311's halo population.

Finally, we analysed the root mean square velocity, rotation measure
and angular momentum parameter as function of galactocentric radius in
Figure\,\ref{fig:vsig_lambda}. For that, as well as for the dynamical modelling
(Section\,\ref{sec:discussion}) we apply a more stringent velocity error
selection, $\sigma(V_{\rm LOS})<60$\,km\,s$^{-1}$, to get a high quality
data sample. Data points close to NGC\,3309 and HCC\,007 are
excluded here as well.

\subsection{Line-of-sight velocity}
\label{sec:res_vel}

The velocity ($V_{\rm LOS}$) map can be characterized by several main 
(bulk motion) regions, but also considerable velocity variations on small
scales. The velocity field within 10\,kpc of NGC\,3311 is rather smooth
with indications of a slight 'asymmetry' in the sense that $V_{\rm LOS}$
is shifted to higher velocities in the North-East than in the South-West
(see lower left panel
in Fig.\,\ref{fig:velsig_profiles}). There is, however no sign of ordered rotation
around NGC\,3311. The system seems to be pressure supported, the
velocity dispersion dominates any potential rotation signal at all radii (see
also Section\,\ref{sec:vsig_lambda}).

\begin{figure}
\centering
\includegraphics[width=0.98\linewidth]{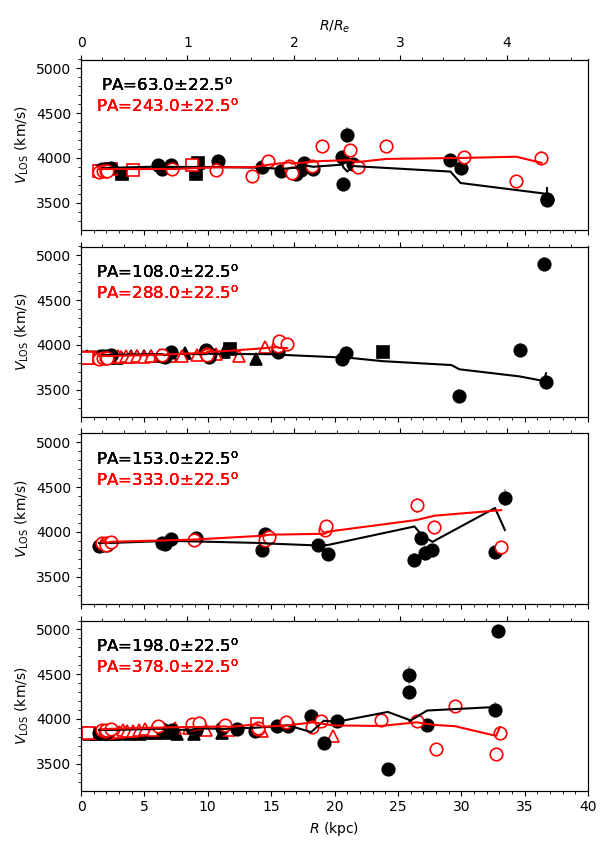}
\caption{Line-of-sight velocity as function of galactocentric radius
for four cones of 45 deg width as shown in Fig.\,\ref{fig:velsig_profiles}.
Black (red) symbols indicate the measured velocities for position angles
lower (larger) than 220.5$^{\rm o}$, and the black (red) solid lines
indicate the smoothed velocity from the LOESS algorithm (see
Fig.\,\ref{fig:kinmaps_loess}. Data points around NGC\,3309 and
HCC\,007 (open orange symbols in Fig.\,\ref{fig:velsig_profiles} were
excluded.}
\label{fig:velocity_folded}
\end{figure}

\begin{figure}
\centering
\includegraphics[width=0.98\linewidth]{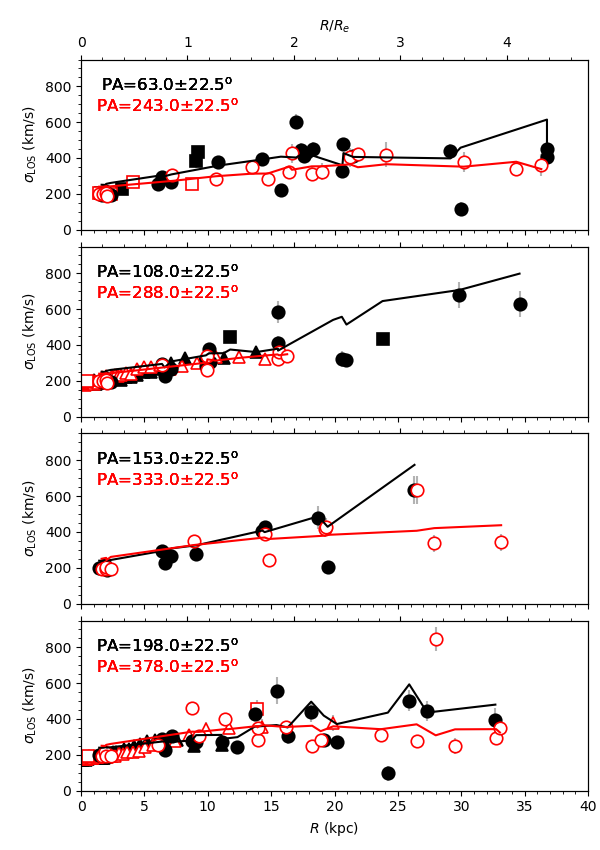}
\caption{Same as figure \ref{fig:velocity_folded} but for the line-of-sight
velocity dispersion.}
\label{fig:dispersion_folded}
\end{figure}

The radial velocities measured in the six most central slits
(cen1\_s27, cen1\_s29, cen2\_s30 and cen2\_s32) are between 3848
and 3879 km\,s$^{-1}$, consistent within the errors with previous
measurements of NGC\,3311's systemic velocity
\citep[e.g.][but NED: 3825 km\,s$^{-1}$]{2011A&A...531A.119R}. We
adopt a central velocity of 3863\,km\,s$^{-1}$ for our  work, which is the
average value of the six central measurements.

Beyond a galactocentric distance of $\sim$10\,kpc from NGC\,3311
the velocity field gets inhomogeneous, that is it varies with position angle.
Regions of high $V_{\rm LOS}$ values are found around NGC\,3309 
($\sim$4095\,km\,s$^{-1}$) and HCC\,007 ($\sim$4760\,km\,s$^{-1}$),
which is expected due to their higher systemic velocities (see open orange
symbols in Fig.\,\ref{fig:velsig_profiles}). It is interesting
to note that around HCC\,007 there seems to exist a velocity gradient
with lower velocities towards NGC\,3311 and larger ones away from it
(see black line bottom panel of Fig.\,\ref{fig:velocity_folded}). The galaxy
HCC\,007 itself shows a clear rotation signal with receding velocities on
its east side and approaching velocities on the West (see Figures
\ref{fig:hcc007_maps} and \ref{fig:hcc007_kinematic}).

The (south-)eastern halo (70$<$PA$<$180$^{\rm o}$) of NGC\,3311
is dominated by low radial velocities (see left middle panels in
Fig.\,\ref{fig:velsig_profiles} and black dots in the middle panels of
Fig.\ref{fig:velocity_folded}). Towards the North-East, in the region of
the excess halo light, velocities scatter quite a lot from slit to slit around
NGC\,3311's systemic velocity (see the left top and bottom panels of
Fig.\,\ref{fig:velsig_profiles}). In general, the
velocities derived from previous long-slit work agree very well with our
velocity map, once systematic offsets to the systemic velocity are
corrected for (see left panels of Figs.\,\ref{fig:velsig_profiles} and
\ref{fig:velsig_pa}). Indeed, the long-slit results
at PA$=64^{\rm o}$ from \citet{2010A&A...520L...9V} show a systematic
offset of $\sim$72\,km\,s$^{-1}$ (3778\,km\,s$^{-1}$ close to the centre)
with respect to our measurements, whereas the central velocity of
$\sim$3864\,km\,s$^{-1}$ measured by \citet{2011A&A...531A.119R}
is close to our measured value. For Figures \ref{fig:kinmaps},
\ref{fig:kinmaps_loess}, \ref{fig:velsig_profiles}, \ref{fig:velsig_pa}
and \ref{fig:velocity_folded} we applied offsets of
$+$72\,km\,s$^{-1}$ and $-$14\,km\,s$^{-1}$ to the long-slit velocities
of Ventimiglia et al. and Richtler et al., respectively.

\subsection{Line-of-sight velocity dispersion}
\label{sec:res_sig}

The velocity dispersion ($\sigma_{\rm LOS}$) of NGC\,3311 rapidly
increases with galactocentric distance from $\sim$185\,km\,s$^{-1}$
at the centre (slit cen2\_s32) to $\sim$500\,km\,s$^{-1}$ at
$\sim$18\,kpc (see Fig.\,\ref{fig:velsig_profiles}). This was already
noted by several previous works 
\citep[e.g.][]{2008MNRAS.391.1009L, 2010A&A...520L...9V, 
2011A&A...531A.119R}. However, the rising $\sigma_{\rm LOS}$ profile
varies for different position angles (see Figures\,\ref{fig:velsig_profiles}
and \ref{fig:dispersion_folded}), meaning that there exist small scale variations
of the local velocity dispersion. This also explains the apparent discrepancies
in the $\sigma_{\rm LOS}$ profiles obtained from the previous long-slit works
\citep[see Figure\,4 in][]{2011A&A...531A.119R}. These differences vanish
when looking at the complete 2-dimensional $\sigma_{\rm LOS}$ distribution.
The long-slit results from Ventimiglia et al. and Richtler et al. are both
consistent with our work.

The regions of highest velocity dispersions are towards the North-East
and east of NGC\,3311 and in the South next to the lenticular galaxy
HCC\,007 (at PA$\sim190^{\rm o}$, see bottom right panel of
Fig.\,\ref{fig:velsig_profiles}).
These regions coincide with the offset envelope and sub-structure from
disrupting dwarf galaxies in the Hydra\,I cluster core, as reported by
\citet{2012A&A...545A..37A}. This is also seen in
Fig.\,\ref{fig:dispersion_folded}, where outside a radius of $\sim$10\,kpc
the average velocity dispersion is higher on the semi-major axis towards
North-East (PA$=$108$^{\rm o}$) than towards South-West
(PA$=$288$^{\rm o}$). The same is true for the directions
East and South-East, although the opposite directions are dominated by
the low-velocity dispersion light of NGC\,3309 (see orange circles in the
right panels of Figures \ref{fig:velsig_profiles} and \ref{fig:velsig_pa}).
The high velocity dispersion east of HCC\,007 is probably caused by the
superposition of the galaxy's stellar population at high radial velocity
(see Fig.\,\ref{fig:hcc007_maps}) with that of NGC\,3311's stellar halo at
lower radial velocities. 

Low velocity dispersion regions are those around the giant elliptical
NGC\,3309 ($\sim$200\,km\,s$^{-1}$) and HCC\,007
($\sim$80\,km\,s$^{-1}$), which dominate the light in those slits.
Curiously, also towards the North-East (0$<$PA$<$45$^{\rm o}$)
there exists a region of relatively low velocity dispersions with
$\sigma\sim$250-280\,km\,s$^{-1}$ (see Figs.\,\ref{fig:kinmaps} and
\ref{fig:kinmaps_loess}. One might speculate whether the light in those
slits is dominated by the pure stellar halo of NGC\,3311 or rather cold
sub-structure in the displaced halo region.

We also calculate the second order moment of velocity as the
root mean square velocity:
\begin{equation}
V_{\rm RMS} = \sqrt{V^2+\sigma^2},
\end{equation}
where $V$ is defined as $|V_{\rm LOS}-V_{\rm NGC\,3311}|$.
Velocity dispersion and $V_{\rm RMS}$ are shown in comparison in
Fig.\,\ref{fig:vsig_lambda} (upper panel). The differences are negligible
below 25\,kpc and still small beyond that radius because $V/\sigma$
is generally low. We also note that, In the case of NGC 3311, $V$
cannot be interpreted as a measure of angular momentum (rotation)
but describes a velocity bias, since the galaxy itself is displaced in
space and velocity from the core of the galaxy cluster 
\citep[e.g.][]{2018A&A...609A..78B}. Thus $V_{\rm RMS}$ cannot be
applied to our simple dynamical modelling. We, therefore, take the
velocity dispersion values for our dynamical analysis in
Sect.\,\ref{sec:disp_models}.

\subsection{Significance of velocity and dispersion scatter}
\label{sec:res_scatter}

In the sections before we claim to see evidence for an intrinsic scatter
in the radial velocities and velocity dispersions distributions for
galactocentric radii $>10$\,kpc.
In order to evaluate the significance of this intrinsic scatter we statistically
compare the distribution of $V_{\rm LOS}$ and $\sigma_{\rm LOS}$
values in the radial range $10<R<40$\,kpc with those of five million
Monte Carlo realisations that are based on measurement uncertainties
only. For this analysis, we use our error selected sample from
Figures\,\ref{fig:velsig_profiles} to \ref{fig:dispersion_folded}
and exclude data points near NGC\,3309 and HCC\,007.

We use the Anderson-Darling test implemented in the python package
\texttt{scipy.stats} based on \cite{Anderson_test} for the comparison. 
This test has been shown to perform better than the usually widely
used Kolmogorov-Smirnov test \citep[e.g][]{Razali2011} since it is
more sensitive to the differences in the wings of the distributions.
We test the null hypothesis that the distribution and scatter of points
is purely caused by measurement errors via this procedure:

Firstly, we define the observed distribution of points as function of
radius as our `reference sample'.
Then, the null hypothesis sample, called `comparison sample',
is created from the reference sample by assigning a de-trended mean
value to each measurement and adding to it randomly generated
gaussian scatter based on its measured uncertainty. The de-trending is
done by an error weighted least square fit to the data.

Secondly, the two distributions are compared via the Anderson-Darling
test. If the observed distribution is caused purely by uncertainties
then the comparison sample should be statistically the same as the
reference sample.
To get the likelihood of how often the comparison sample 
resembles the observed distribution we generate the comparison
sample five million times and obtain $p$-value distribution of the
Anderson-Darling tests. We compute the probability of the $p$-value
being smaller or equal to $0.05$, which is $5\%$ significance level
that is usually assumed as a statistically significant difference between
two distributions. This takes into account the small number statistics
and the observation uncertainties.

Finally, we compute the significance of the deviation in terms of
sigma, assuming a gaussian distribution. 

\begin{figure}
\centering
\includegraphics[width=0.98\linewidth]{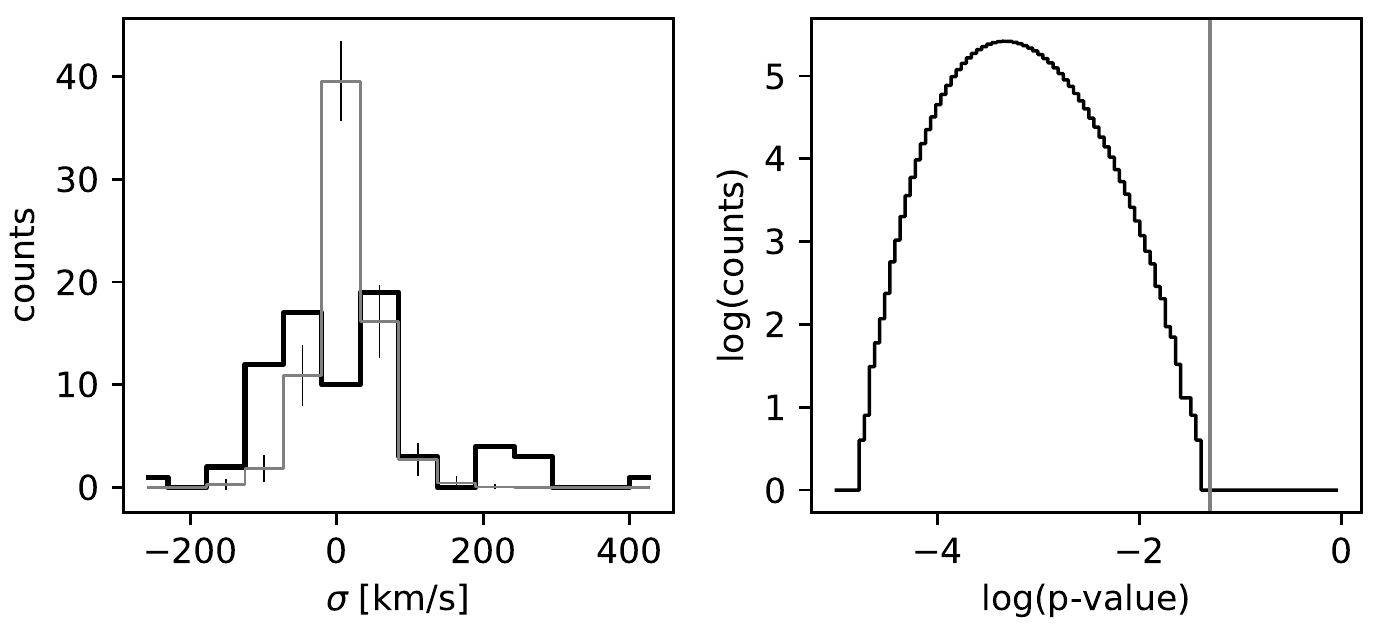}
\caption{Left: Distribution of observed, but trending corrected,
velocity dispersions between 10 to 40\,kpc from NGC\,3311 (thick
black line) compared to the average distribution of five million
realisations of mock distributions that consider the observational
errors only (grey thin line).
Right: Distribution of five million $p$-values from the
Anderson-Darling test, with the vertical grey line indicating the 
$5\%$ significance level. The significance of an intrinsic scatter
in the observed $\sigma_{\rm LOS}$ distribution is 4.8 $\sigma$.}
\label{fig:disp_scatter}
\end{figure}

In Fig.\,\ref{fig:disp_scatter} we show the results of our tests for the
distribution of velocity dispersions in the radial range 10 to 40\,kpc
from NGC\,3311. The left plot shows the distribution of the de-trended
reference (observed) sample and the average distribution of five million
realisations of the (simulated) comparison sample, The right plot
shows the distribution of five million $p$-values from the
Anderson-Darling test, with the vertical line indicating the  $5\%$
significance level. The low probabilities that both distributions are
equal translate into a significance of 5.1 $\sigma$ that the observed
scatter of the $\sigma_{\rm LOS}$ distribution is intrinsic and cannot
be explained by the observational errors alone.
Restricting the radial range to 10-20, 10-30 or 15-40\,kpc also results
in significant intrinsic velocity dispersion scatter with $\sigma$-values
of  3.4, 4.6 and 3.3, respectively. 

Analogous tests for the distribution of radial velocities lead to a
4.8 $\sigma$ significance of an intrinsic scatter in the velocity
distribution. Restricting the velocity sample to the east side of 
NGC\,3311, the region that is not influenced by NGC\,3309 and
HCC\,007, still results in a significance of 3.1 $\sigma$, despite
a lower number statistics of data points.

\subsection{Higher moments of the line-of-sight velocity distribution}
\label{sec:res_h3h4}

The higher moments $h_3$ (skewness) and $h_4$ (kurtosis) of the
line-of-sight velocity distribution are useful quantities in the case of a
relaxed dynamical system. In a pressure supported system with
negligible rotation and no orbital anisotropies both moments should
have values around zero.

As can be seen in Figures \ref{fig:kinmaps_loess} and
\ref{fig:h3h4_profiles} $h_3$ scatters mostly around zero, in particular
in the central 10\,kpc around NGC\,3311. In the north-east region of
the displaced halo $h_3$ tends to scatter to negative values (see left top
and bottom panels of Fig.\,\ref{fig:h3h4_profiles} for $>$20\,kpc and
$<-$20\,kpc, respectively). It is also mostly negative around NGC\,3309
(see left second panel from top in Fig.\,\ref{fig:h3h4_profiles}, negative
spatial axis). That means that the LOSVD has
an extended tail towards lower velocities. In the case of NGC\,3309
this is probably due to a contribution of NGC\,3311's halo light to the
dominating stellar populations of NGC\,3309 itself, which has a higher
radial LOS velocity than NGC\,3311. In the displaced halo region and
east of NGC\,3311, radial velocities scatter towards lower values than
the systemic velocity of NGC\,3311 (see Fig.\,\ref{fig:velsig_profiles}),
in agreement with a negative $h_3$ value. Part of the displaced halo
light and dwarf galaxy tidal tails seem to be dominated by low-radial
velocity stellar populations.

The moment $h_4$ is, with a few exceptions, always positive (see
Figures \ref{fig:kinmaps_loess} and \ref{fig:h3h4_profiles}). At face
value this points to radially biased orbits all around NGC\,3311.
However, this interpretation is most probably not correct, since there
is a lot of sub-structure in the stellar halo of NGC\,3311. Thus, we
might rather see the superposition of LOSVDs from these different  
sub-structures, as we have also argued in
\citet{2018A&A...609A..78B}.

The $h_3$ and $h_4$ values around HCC\,007 are interesting. They
show the typical pattern of negative to positive $h_3$ values on both
sides of a rotating galaxy (see Fig.\,\ref{fig:hcc007_maps}). Even in
the wider environment $h_3$ seems to show a gradient from negative
to positive values with increasing distance from NGC\,3311 (see lower
left panel in Fig.\,\ref{fig:h3h4_profiles}).

\subsection{Rotation measure and angular momentum parameter}
\label{sec:vsig_lambda}

The rotation measure $V/\sigma$ and the projected specific angular
momentum parameter $\lambda_R$ \citep{2007MNRAS.379..401E}
are useful quantities to examine the dynamical status of a galaxy and 
to kinematically classify early-type galaxies 
\citep[e.g.][]{2007MNRAS.379..418C}.

\begin{figure}
\centering
\includegraphics[width=0.98\linewidth]{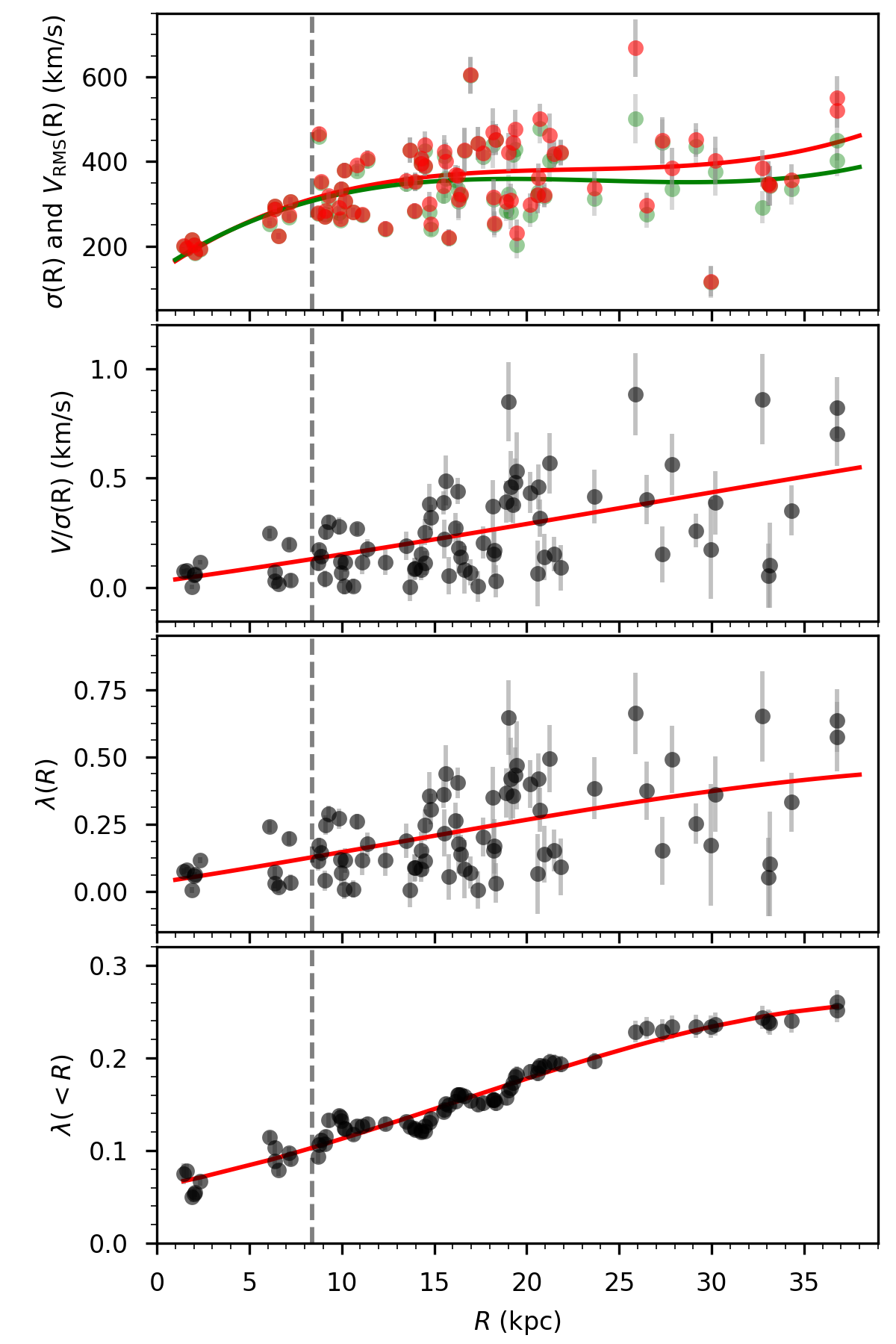}
\caption{Top: Velocity dispersion (green) and root mean square
velocity, $V_{\rm RMS}$ (red) as function of galactocentric distance to
NGC\,3311. Middle top: Rotation measure $V/\sigma$. Middle bottom:
Local angular momentum parameter  $\lambda(R)$.
Bottom: Cumulative angular momentum parameter $\lambda(<R)$
within a galactocentric distance $R$. In all four panels the vertical
dashed line indicates the effective radius of NGC\,3311. The blue,
green and red curves show polynomial fits of third order to the data.
Data points close to NGC\,3309 and HCC\,007 were excluded from
the analysis.}
\label{fig:vsig_lambda}
\end{figure}

In our case, we calculate $V/\sigma$ for each Voronoi bin/slit
individually and evaluate its trend with galactocentric distance to
NGC\,3311. $V$ is defined as $|V_{\rm LOS}-V_{\rm NGC\,3311}|$
and $\sigma$ is the average of the line-of-sight velocity dispersion in
the respective bin.

For $\lambda_R$, we calculate the local angular momentum parameter
$\lambda(R)$ as
\begin{equation}
\lambda(R) = \frac{R V}{R  \sqrt{V^2+\sigma^2}},
\end{equation}
and the cumulative angular momentum parameter $\lambda(<R)$ within
a certain radius $R$ around NGC\,3311 as
\begin{equation}
\lambda(<R) = \frac{\sum_{n=1}^{N(<R)} F_n\, R_n |V_n|}
{\sum_{n=1}^{N(<R)} F_n\,R_n  \sqrt{V_n^2+\sigma_n^2}},
\label{eq:lambda}
\end{equation}
where the data are ordered with increasing projected distance to
NGC\,3311 and the flux $F_n$ is approximated by the $S/N$ value
of the spectrum in the corresponding slit (see Table\,\ref{tab:kinematics}).

The results are shown in Fig.\,\ref{fig:vsig_lambda} where we plot
$V/\sigma$, $\lambda(R)$ and $\lambda(<R)$ (from top to bottom)
as function of galactocentric distance from NGC\,3311. We excluded
from this plot (and the calculation of $\lambda(<R)$) the slits around
NGC\,3309 and those of HCC\,007 (open orange symbols in
Fig.\,\ref{fig:velsig_profiles}), since we are only interested in
the dynamical properties of the central galaxy. Also, we restricted
our dataset to measurements with a velocity dispersion error lower
than 60\,km\,s$^{-1}$.

At basically all radii the stellar halo of NGC\,3311 is dispersion
dominated ($V/\sigma<1$). Within the effective radius of 8.4\,kpc
(vertical dashed line), $V/\sigma<0.2$, beyond this point it shows a
larger scatter and its average value (red curve) increases to
$V/\sigma\simeq0.5$, on average, at 35\,kpc distance.
\citet{2013MNRAS.428..389P} reported  $V/\sigma=0.03\pm0.14$ for
NGC\,3311, derived from the kinematics of 116 ultra-compact dwarf
galaxies and bright globular clusters from \citet{2011A&A...531A...4M}.
This low value is in general agreement with the stellar halo data,
although the small sample of very massive GCs (some of them
probably being UCDs) might give a rather biased view of the typical
halo population around NGC\,3311.

$\lambda(R)$ shows a similar behaviour as $V/\sigma$, with an
increasing scatter at larger distances. The cumulative global angular
momentum $\lambda(<R)$ is very low within the effective radius.
At $R_{\rm eff}$ it reaches a value of $\lambda_{R_{\rm e}}=0.1$.
This is in the regime of slow rotators \citep{2007MNRAS.379..401E}
and typical for central cluster galaxies with a small ellipticity 
\citep[$\varepsilon_{\rm NGC\,3311}=0.05$][]{2012A&A...545A..37A}.
The steady increase of $\lambda(<R)$ to 0.26 at 35\,kpc shows
that there might be some angular momentum present in the outskirts
of NGC\,3311. We will discuss this in the following section.

\subsection{Comparison with other central cluster galaxies}

Searching the literature for similarities with NGC\,3311, one searches
for a nearby and rather low-mass central galaxy with  two-dimensional
velocity field which shows a strongly rising velocity dispersion profile.
While some mild counterparts may be identified, very similar objects
are rare.
 
The MASSIVE survey targets the $\sim$100 most massive early-type
galaxies ($M\geq10^{11.5}M_{\odot}$) within a distance of 108\,Mpc
with IFU spectroscopy \citep{2014ApJ...795..158M}. 65 of them are
brightest group galaxies (BGGs) and a few are brightest cluster galaxies
(BCGs).
\citet{2017MNRAS.464..356V} presented the kinematics of 41 galaxies
from the MASSIVE survey. 80\% of them are slow or non-rotators.
The 12 most luminous galaxies in that sample are BGGs.
Out to 20-25\,kpc (or 2-3 effective radii), they show
rising or nearly flat velocity dispersion profiles from central values of
$\sim240$\,km\,s$^{-1}$ to $\sim350$\,km\,s$^{-1}$, similar to the
values of NGC\,3311 at the same distance. But it is is not clear what is
their kinematical behaviour outside 2-3 effective radii, where the
velocity dispersion profile of NGC\,3331 is rising to the level of
500\,km\,s$^{-1}$ in its outer halo.
Galaxies with a mildly rising velocity dispersion profiles are NGC\,1129
and NGC\,7242. However, it is striking that some galaxies show a large
scatter in their dispersion profile (NGC\,3158, NGC\,507, NGC\,1016),
indicating that Jeans models with one Jeans equation only (unique
density profile, anisotropy, etc.) may not be appropriate.

Also other kinematical studies of massive galaxies are consistent with
the occurrence of flat and rising velocity dispersion profiles
\citep{1999MNRAS.307..131C, 2002ApJ...576..720K,
2008MNRAS.391.1009L, 2009MNRAS.394.1249C,
2013MNRAS.428..389P, 2016MNRAS.457.1242F}. 
\citet{2008MNRAS.391.1009L} presented radial profiles for the 
line-of-sight velocity and velocity dispersion from long-slit observations
of 41 BCGs, most of which they classify as dispersion supported, and
which also show a variety of dispersion profile shapes, with a fair
fraction of galaxies with flat or rising dispersion profiles, although not
with the same amplitude of $\sim$600\,km\,s$^{-1}$ as for NGC\,3311.

\citet{2011ApJ...728L..39N} find the BCG in A383 to exhibit a
steeply rising dispersion profile that climbs from $\sim$270\,km\,s$^{-1}$
at the centre of the galaxy to $\sim$500\,km\,s$^{-1}$ $\sim$ 22\,kpc,
a similar behaviour as for NGC\,3311.
\citet{2016MNRAS.457.1242F} compared 24 early-type galaxies from
the SLUGGS survey to the assembly classes of galaxy simulations by
\citet[][see also Sect.\,\ref{sec:res_simul}]{2014MNRAS.444.3357N}.
They assign 6 galaxies to the classes C, E and F, which are the most
appropriate to describe a massive, central slow rotating galaxy.
In therms of the high-order moments $h3$ and $h4$, the kinematic
properties of NGC\,5846, a central group galaxy, seems to show most
similarity to NGC\,3311, followed by NGC\,4374 and 4365, although their 
velocity dispersion profiles are not rising but are rather flat.

Positive $h_4$ values at all radii have also been found for the three
BCGs NGC\,6166, 6173 and 6086 in the study of
\citet{1999MNRAS.307..131C} .
Also most galaxies in the MASSIVE survey show positive average
$h_4$ values \citep{2017MNRAS.464..356V}.
Positive $h_4$ values generally indicate a bias towards radial orbits
\citep{1993MNRAS.265..213G, 1997ApJ...488..702R,
1998MNRAS.295..197G}.
Since $h_4$ stays nearly constant at a value of $\sim0.1$ at all radii,
the increase in velocity dispersion is not associated with a change in
velocity anisotropy towards tangential orbits. It rather reflects an
increasing mass-to-light ratio, and thus a massive dark halo.

The lack of a significant rotation signal for NGC\,3311 is consistent
with the view that most BCGs are slow rotators, mostly due to their
build-up from multiple (dry) mergers. In a VIMOS IFU study of 
10 BCGs at redshift $z=0.1$ with masses
$10^{10.5}<M_{\rm dyn}<10^{11.9}M_{\odot}$ 
\citet{2013ApJ...778..171J} found that 70\% are slow rotators,
and above $M_{\rm dyn}\sim10^{11.5}M_{\odot}$ all BCGs are slow
rotators. 
The steadily increasing cumulative angular momentum profile of
NGC\,3311 lies above most radial profiles of BCGs from the
MASSIVE survey \citep{2017MNRAS.464..356V}.

The probably two best studied cases of central galaxies with rising
velocity dispersion profiles in their outer halo that are similar to 
NGC\,3311 are M87 in the Virgo cluster and NGC\,6166 in Abell\,2199.
\citet{2014ApJ...785..143M} measured the line-of-sight velocity
distribution from integrated stellar light at two points in the outer halo
of M87. They found a rising velocity dispersion up to 577\,km\,s$^{-1}$,
but argue that there is evidence for two kinematically distinct stellar
components. They conclude that the asymmetry seen in the velocity
profiles suggests that the stellar halo of M87 is not in a relaxed state
and complicates a clean dynamical interpretation.
For NGC\,6166 we predict a similar complex velocity dispersion field as
found for NGC\,3311.

\subsection{Results in context of cosmological simulations}
\label{sec:res_simul}

The growth of massive early-type galaxies appears to happen in two 
main phases. The early mass assembly (2$<z<$6) is dominated by
significant gas inflows \citep[e.g.][]{2005MNRAS.363....2K,
2009Natur.457..451D} and the in situ formation of stars. The late
evolution is dominated by the assembly of stars which have formed
in other galaxies and have then been accreted via interactions or
mergers on to the central galaxy at lower redshifts (0$<z<$1)
\citep[e.g.][]{2003ApJ...590..619M, 2007ApJ...658..710N}.

The cosmological simulations of this two-phase galaxy formation
process by \citet{2010ApJ...725.2312O} predict that the present day
masses of very massive galaxies is dominated at the level of 80\%
by accretion and merging since redshift $z\leq3$.
Also \citet{2016MNRAS.458.2371R} showed, using the Illustris
simulations, that 80\% of the stars in very massive galaxies
($M_{\ast}\sim10^{12}M_{\odot}$) are born ex situ and then
were accreted onto the galaxies via mergers.
This means that the outer stellar dynamics of a central massive
galaxy are dominated by the orbits of accreted material.
In particular, minor mergers of galaxies embedded in massive dark
matter halos provide a mechanism for explaining radially biased
velocity dispersions at large radii \citep{2012MNRAS.425.3119H}.
Also cosmic zoom simulations of massive galaxies
\citep{2014MNRAS.438.2701W} show that massive galaxies with a
large fraction of accreted stars have radially anisotropic velocity
distributions outside the effective radius.

\citet{2014MNRAS.444.3357N} presented a
two-dimensional dynamical analysis of a sample of 44 cosmological
hydrodynamical simulations of individual central galaxies up to
6$\times$10$^{11}M_{\odot}$. They defined the classes A-F to
characterise the different assembly histories. NGC\,3311 is different
than most types. If any, type C or F is closest to the properties of
NGC\,3311 (see Table 2 of Naab et al.): map feature is a dispersion
dip, no correlation of $h_3$ with $v/\sigma$.

\citet{2017MNRAS.467.4101G} investigated the mass growth of
BCGs since redshift $z<0.3$. They found an increasing pair fraction,
with decreasing redshift, which they interpret as an increasing merger
fraction. The mass growth via accretion from close pairs is $24\pm14$\%
since $z=0.3$. Most of this mass is deposited into the intracluster light.
NGC\,3311 has the close pair NGC\,3309. Although no direct
interaction signs are seen in our data, it is plausible that some outer
material of NGC\,3309 has already been stripped off in the cluster
potential and might contribute to the central cD halo in the future.

\citet{2017ApJ...841...45Y} used the Illustris simulations to investigate
the displacement and velocity bias of central galaxies with respect to their
dark matter haloes. The central galaxy velocity bias can be explained by
the close interactions between the central and satellite galaxies. A central
velocity bias naturally leads to a small offset between the position of the
central galaxy and the halo potential minimum \citep{2015MNRAS.446..578G}.
This scenario might explain the offset halo we see around NGC\,3311.

\section{Discussion}
\label{sec:discussion}

In Paper\,I, where we analysed the metallicity, [$\alpha$/Fe] abundances
and ages of the stellar halo light around NGC\,3311, we concluded
that our findings can be explained by a two-phase galaxy formation
scenario \citep[e.g.][]{2010ApJ...725.2312O}. The inner spheroid
component (within 10\,kpc) was probably
formed in-situ in a rapid collapse very early-on, whereas the outer
halo was accreted over a long timescale, a process that is still
ongoing. Do we also see this in the kinematical features of the galaxy?
Another important question is whether Jeans models are appropriate for
constraining the mass profile when phase space is clumpy. 

\subsection{The velocity dispersion profile from stars to galaxies}

\citet{2011A&A...531A.119R} presented velocity dispersions of bright
globular clusters around NGC\,3311 and, in conjunction with long-slit data
of the central galaxy population, constructed  spherical Jeans models.
Fig.\,4 in Richtler et al. indicates a rising velocity dispersion, which may be 
understood as a class attribute of central cluster galaxies. 
A spherical Jeans model in conjunction with a unique tracer population 
demands a massive dark halo with a large core in order to explain the
rising dispersion profile.

\begin{table}[ht!]
\centering
\label{tab:dispersions}
\caption{Projected velocity dispersions for the stellar population (s), globular
clusters (gc), and galaxies (gal).}
\begin{tabular}{lrrrr}
\hline
\hline
Pop. & Distance & $\sigma_{\rm{LOS}}$ & $\Delta\sigma_{\rm{LOS}}$ &
 N$_{\rm gal}$  \\
  & (kpc) & (km\,s$^{-1}$) & (km\,s$^{-1}$) & \\
(1) & (2) & (3) & (4) & (5) \\
\hline
 s  &    4.04 &      267 &  20 & \\
 s  &   5.12  &     288 & 20  &  \\
 s  &   7.52  &     335 &  20 & \\
 s  &   10.50 &       380 & 20 & \\
 s  &   11.61 &      407 & 20 & \\
 s  &    12.90 &      416 & 20 & \\
 s  &   14.30 &      436 & 20 & \\
 s  &   16.90 &      454 & 20 & \\
 s  &   18.50 &      470 & 20 & \\
 s  &   20.20 &      484 & 20 & \\
 s  &   21.50 &      495 & 20 & \\
 s  &    23.80 &      515&  20 & \\
 s  &   25.67 &      524 & 20 & \\
 s  &   28.07 &      540 & 20 & \\
 s  &   30.12 &       551 & 20 & \\
 s  &   31.92 &      554 & 20 & \\
 gc &    13.900    &   535  &  70 & \\
 gc &    31.3    &   567  &   71 & \\
 gc &    38.30    &  650  &  77 & \\
 gc &    52.20    &  732  &  97 & \\
 gc &    55.70    & 707   &  89 & \\
 gc &    97.50    & 807   &  107 & \\
 gc &    195.00     &   703  &   94 & \\
 gal &    150.00  &   697 & 79 & 39\\
 gal &    250.00  & 798 & 122 & 27\\
 gal &    350.00 &  664 & 73 & 41\\
 gal &    450.00 &  699 & 90 & 30\\
 gal &    550.00 &  725 & 82 & 43\\
 gal &    650.00 &  603 & 83 & 27\\
 gal &    750.00 &  1050 & 200 & 15\\
\hline
\end{tabular}
\end{table}

\begin{figure}
\centering
\includegraphics[width=0.98\linewidth]{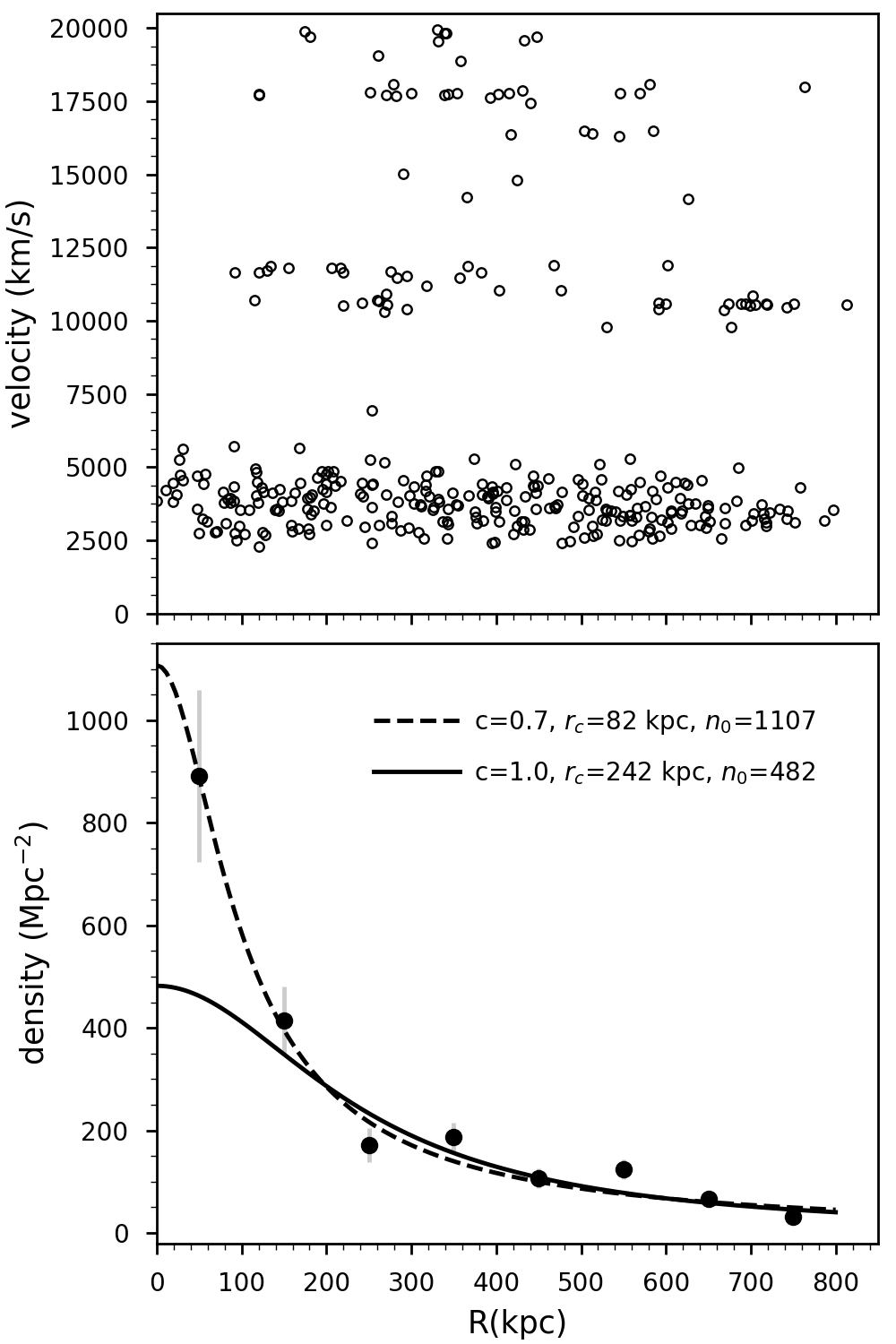}
\caption{The upper panel shows all galaxies within a distance of one degree
from NGC\,3311 with a radial velocity limit of 20000\,km\,s$^{-1}$. The Hydra\,I
members stand out clearly. We select galaxies with a radial velocities less than 
6000\,km\,s$^{-1}$ km/s. The lower panel shows the surface number densities
in 8 bins. The parameters of two modified Hubble profiles are given, one of
them fitted with the first bin skipped. The core radius is only badly defined and
may be high.}
\label{fig:gal_densities}
\end{figure}

Because we want to discuss our results in the context of the entire cluster,
including the galaxy population, Table\,\ref{tab:dispersions} lists the relevant
velocity dispersion values for the inner stellar population, the globular
clusters, and the cluster galaxies. The first column indicates the population,
the second the cluster-centric distance, the third and forth the velocity
dispersion and its uncertainty, and the fifth column with the number of 
galaxies in the bin, as detailed below.

The quoted values for stars and globular clusters come from 
\citet{2011A&A...531A.119R}. The values for the galaxies have been derived
in the following manner: we selected from the NASA Extragalactic Database
all objects classified as galaxies within a radius of one degree around
NGC\,3311, corresponding to 885\,kpc (there are hardly any objects outside),
and end up with 638 galaxies. These galaxies are nicely stratified in redshift.
The Hydra\,I members stand out clearly as the upper panel of
Fig.\ref{fig:gal_densities} shows (for readability only until 20000\,km\,s$^{-1}$
as the maximal radial velocity). We select Hydra\,I members as having radial
velocities less than 6000\,km\,s$^{-1}$. This sample comprises 251 galaxies. 
We build 8 bins of widths 100\,km\,s$^{-1}$ and calculate the surface number
density as shown in the lower panel of Fig.\,\ref{fig:gal_densities}. 

The surface number densities are fit by the modified Hubble profile
\begin{equation}
n(r)  = \frac{n_0}{[1 + (r/r_c)^2]^{-c} }
\end{equation} 
with $r_c$ as the core radius. Two fits, one skipping the innermost bin,
are indicated. The important point is that the core radius is not well
constrained. At larger cluster-centric distance it may assume values even
above 300\,kpc. This must be remembered when constructing Jeans models,
because the core radius is related to the difference in the gravitational
potential experienced by a tracer population: a small core radius means a
small potential difference, a large core radius a large potential difference.
For each bin we calculate the velocity dispersion applying the maximum
likelihood velocity dispersion estimator and its uncertainty given by
\citet{1993ASPC...50.....D}. 

\subsection{A MONDian model}

The mass profiles of galaxy clusters argue for the existence of cluster dark
matter even with the validity of Modified Newtonian Dynamics (MOND)
\citep[e.g.][]{2015MNRAS.454.3810M} and it is clear that a rising velocity
dispersion in NGC\,3311 strongly contradicts MONDian predictions at first sight.
However, we find it interesting where MOND appears in our context.
Therefore we briefly summarise how we calculate the MONDian circular
velocity. Our technique is described, for example, for the case of NGC\,4636
by \citet{2006A&A...459..391S, 2012A&A...544A.115S} and we refer the
interested reader to these works. 

Significant MONDian effects are expected, when the acceleration,
$g = V_{\rm circ}^2/R$, is smaller than the  constant 
$a_0 = 1.35\times10^{-8} {\rm cm/s^2}$ ($V_{\rm circ}$ is the circular velocity
and $R$ the galactocentric radius). The recipe for calculating the MONDian
circular velocity $g_M$  from the Newtonian circular velocity $g_M$ in case
of sphericity is 
\begin{equation}
g_N = \mu(g/a_0) g_M
\end{equation}
where the function $\mu$ interpolates between the Newtonian and the
MONDian regime. We chose the `simple' formula, $\mu =  x/(x+1)$, with
$ x= g_M/a_0$.

We consider the baryonic mass profile as the sum of the stellar NGC\,3311
mass profile and the X-ray mass after \citet{2006PASJ...58..695H} which
we compute as
\begin{equation}
M_x (r) = 4\pi\times n_0 \times (r_c^2 r + r_c^3 \arctan(r/r_c))
\end{equation}
with $n_0 = 1.59\times10^{-4} M_\odot/{\rm pc}^3$ and $r_c =$102\,kpc.
The MONDian circular velocity then follows as 
\begin{equation} 
v_{M} = \sqrt{v_N^2(r)/2 + \sqrt{v_N^4(r)/4 + v_N^2(r) a_0 r}}.
\end{equation}
The difference  between the purely baryonic mass and the MONDian mass
is called the `MONDian ghost halo'.  The  projected velocity dispersions
are then calculated as in any Jeans model.

\subsection{The velocity dispersions and comparison with models}
\label{sec:disp_models}

\begin{figure}[t]
\centering
\includegraphics[width=0.98\linewidth]{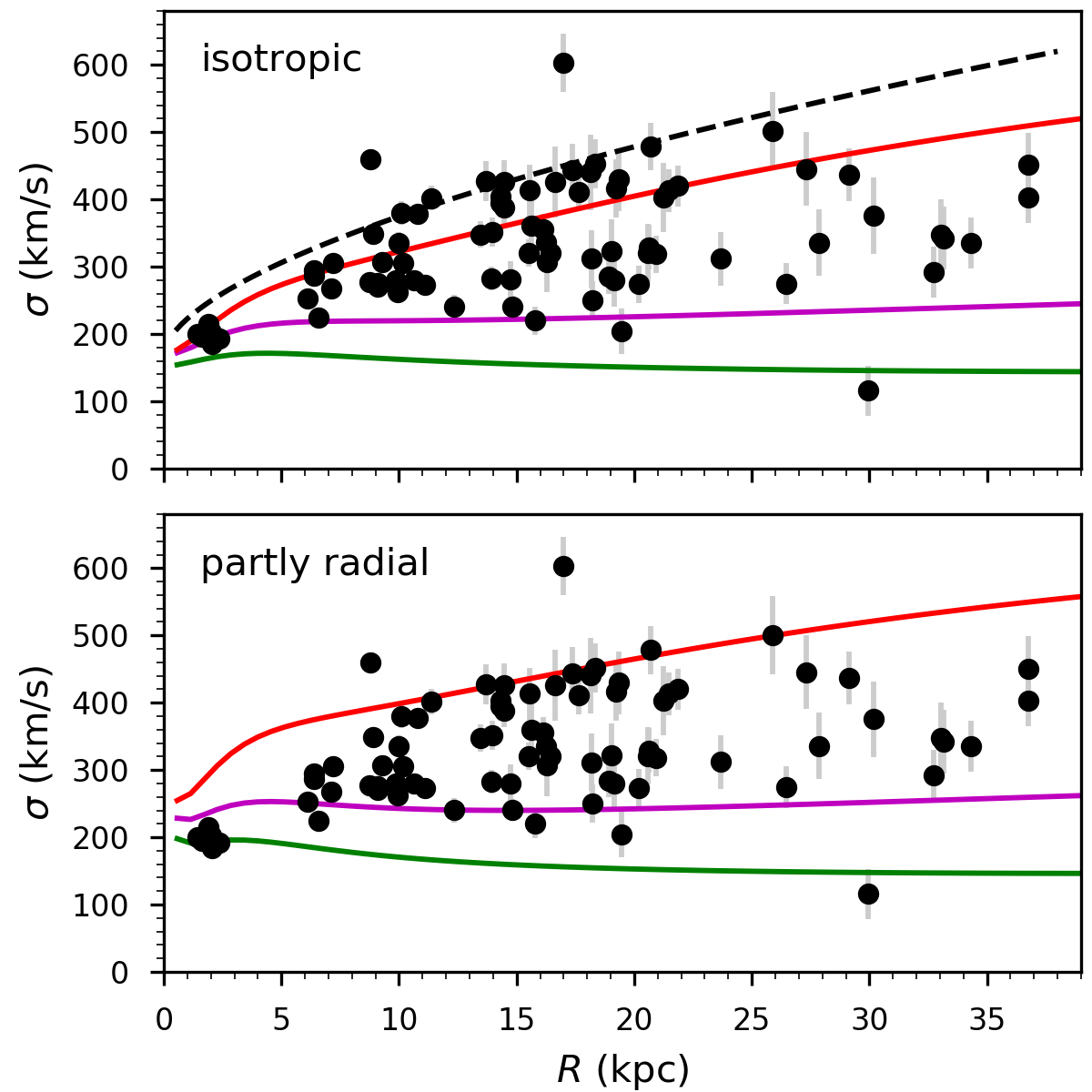}
\caption{Spherical Jeans models for the line-of-sight velocity dispersion
of our NGC\,3311 halo population data (with flag=1 in
Table\,\ref{tab:kinematics}) as function of galactocentric distance to
NGC\,3311 for the isotropic (top panel) and partly radial (bottom panel) case.
The green curve shows the baryonic component only (stellar mass with
$M/L_R=5.0$ and mass of X-ray gas). The magenta curve is a MOND
model. The red curve is the sum of the baryonic mass (green) and a Burkert
halo. See text for more details on the models. In the upper panel the
dashed line is an approximate fit to the upper envelope of the distribution.}
\label{fig:dispersion_models1}
\end{figure}

\begin{figure}[t]
\centering
\includegraphics[width=0.98\linewidth]{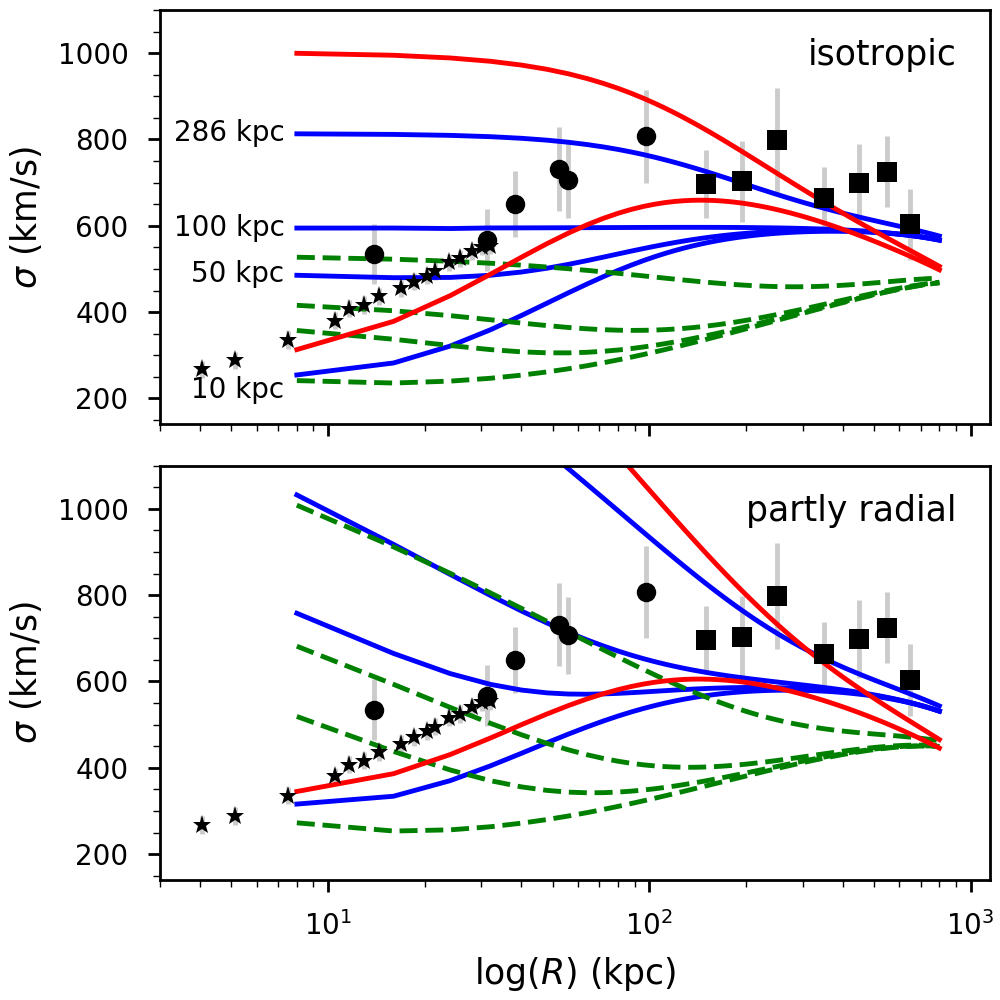}
\caption{Spherical Jeans models for the line-of-sight velocity dispersion
of several tracer populations as function of galactocentric distance to
NGC\,3311 for the isotropic (top panel) and partly radial (bottom panel)
case and different core radii as indicated (blue lines). The asterisks mark
the upper envelope of the stellar light
$\sigma_{\rm LOS}$ distribution from our data (dashed line in
Fig.\,\ref{fig:dispersion_models1}). Filled circles represent the velocity
dispersions of globular clusters, and squares the velocity dispersion of
Hydra\,I cluster galaxies. The dashed
(green) lines are MOND models. For comparison, we also show as red
lines the cored Burkert halos from \citet{2011A&A...531A.119R} for the 
extreme core radii.}
\label{fig:dispersion_models2}
\end{figure}

We want to argue with the help of Fig.\,\ref{fig:dispersion_models1}
that the inference of a cored dark matter profile may provide only a partial
description of the dynamical state of the system. The velocity field clearly
depends on position angle, but the one-dimensional presentation of the
profile shows a pattern that is easily overlooked in two-dimensional maps.
The description of our measurements
\citep[including][]{2011A&A...531A.119R} is: within the inner 5\,kpc, the
dispersion rises from a central value of 170\,km\,s$^{-1}$ to 250\,km\,s$^{-1}$.
That can be explained by the baryonic mass profile in conjunction with a small
radial anisotropy. At radii larger than 5\,kpc, the velocity dispersion is no
longer a unique function of radius, but  spreads out in some dispersion
interval whose size increases with radius. This feature is already
insinuated in Fig.\,4 of \citet{2011A&A...531A.119R}. One notes that the
upper and lower envelopes of this distribution are quite well defined.
The lower envelope is constant at a velocity dispersion of 200\,km\,s$^{-1}$
and it traces remarkably well the MONDian prediction (see magenta curve in
Fig.\,\ref{fig:dispersion_models1}), with  few exceptions.
The upper envelope follows ($r$ in kpc),
$\sigma_{\rm upper}$[km\,s$^{-1}$]$=170\times(1+ (r/r_c)^{0.59})$, with
$r_c=7.3$\,kpc (dashed line in Fig.\,\ref{fig:dispersion_models1}).
This also fits well the globular cluster velocity dispersions out to 100\,kpc
(see Fig.\,\ref{fig:dispersion_models2}), which reaches maximum values
of $\sim$800\,km\,s$^{-1}$. In both Figures
(\ref{fig:dispersion_models1} and \ref{fig:dispersion_models2}) only
data points with a velocity dispersion error smaller than 60\,km\,s$^{-1}$
have been considered.

Our central claim is that this scattered distribution of the velocity dispersion
values can be explained by a large range of mass profiles, depending on
the properties of the tracer populations.  
Fig.\,\ref{fig:dispersion_models1} (small scale) and
Fig.\,\ref{fig:dispersion_models2} (large scale) illustrate this claim by
showing two extreme models.
Plotted  are Jeans models using the stellar mass profile from 
\citet{2011A&A...531A.119R}. The surface brightness profile is  
\begin{equation}
\label{eq:light1}
\mu_V(R)=-2.5\log \left(a_1 \left(1+\left( \frac{R}{r_1} \right)^2 \right)^{\alpha_1}
 + a_2 \left (1+\left(\frac{R}{r_2}\right)^2 \right)^{\alpha_2} \right)
\end{equation}
with, $a_1$\,=\,2.602$\times$10$^{-8}$, $a_2$\,=\,1.768$\times$10$^{-9}$,
$r_1$\,=\,5\arcsec, $r_2$\,=\,50\arcsec, $\alpha_1 = \alpha_2$\,=\,$-1.0$.  
This has to be multiplied with the stellar $M/L_R$ ratio. Here we adopt
5.0 instead of 6.0 which better represents the central velocity dispersion.
In the upper panel isotropic models are included while in the lower panel
radially biased models are shown, where we adopt the anisotropy to be  
\begin{equation}
\beta(r) = \frac{1}{2}\frac{r}{r + r_s}
\end{equation}
\citep[formula (60) of][]{2005MNRAS.363..705M} with some small $r_s$
so that practically $\beta = 0.5$ over the entire radius range.

Model\,1 (red curve) is the sum of the stellar mass and the
Burkert-halo of dark matter whose density profile is
\begin{equation}
\rho(r) = \frac{\rho_0}{(1+r/r_c)(1+(r/r_c)^2)}
\end{equation}
with $\rho_0$ being the central density and $r_c$ a characteristic radius.

As a second reference model, we choose a model for MOdified Newtonian
Dynamics (MOND) (magenta curve)
\citep[e.g.][]{1984ApJ...287..571M,2014MNRAS.437.2531M}.
The favoured mass model for NGC\,3311 in \citet{2011A&A...531A.119R} is
clearly non-MONDian, but the new view onto the phenomenon on a rising
velocity dispersion may not longer permit to see NGC\,3311 as a clear-cut
example for a galaxy that needs dark matter beyond the MONDian ghost
halo. 
 
Model 3 (green curve) is the stellar mass only, which we give for
comparison.

\subsection{Meaning of the velocity dispersion pattern}
\label{sec:disp_meaning}

Within a spherical Jeans model, each coordinate communicates with
all other coordinates by an infinite number of tracers following a uniform
density profile. Here we have a different situation in that we obviously
observe the transition from a small scale to a large scale potential that means
different populations with different dynamical histories, their superposition
and their projection. This transition will not be sudden, but smoothed out
during the entire history of infall processes \citep[e.g. see the properties of
the phase space for BCGs in the simulations by][]{2010MNRAS.405.1544D}.

At a given radius, the minimum velocity dispersion must describe an
in situ population whose dynamics is not influenced by the cluster
potential, but by the local distribution of mass, where contributions come
from orbits within about 10-20\,kpc.
Quickly overlapping and becoming visible outwards of radius $r=$10\,kpc,
one finds populations whose dynamical histories are progressively stronger
linked to the large-scale cluster potential. The highest dispersion values
must be caused by the largest potential differences and naturally elongated
orbits. Because of the projection of radially varying fractions of different
sub-populations (e.g. galactic stars, cD halo stars, intra-cluster stars) one
also finds velocity dispersion values in between the minimum and the
maximum dispersion values.

From these considerations, one would conclude that mass determinations
based on Jeans-models with a single tracer population are in our case
not appropriate. The actually observed projected line-of-sight velocity 
distribution is a superposition of several, maybe many, populations with
different samplings of the galaxy and cluster potential, disturbing a 
`clean Jeans-world'.

\subsection{An illustration for the scatter of velocity dispersions}
\label{sec:scatter}

Given the considerations above, the dark matter halo of
\citet{2011A&A...531A.119R} is an ingredient to fullfil the
Jeans-equation for one given population.

We assume that the projected density profiles of the inner stellar population
and the cluster galaxies as the extreme populations can both be described
by $\rho \propto \left[(1+(r/r_c)^2\right]^{-1}$, where $r_c$ is the core radius.
The size of $r_c$ then indicates the different potentials the populations are
living in: a large core radius means that many objects with orbits with
large `apocluster' distances are sampled, a small core radius means
that mainly the `local' population is contributing. The core radius for the
stellar population is 1.1\,kpc. On the other hand, for the cluster galaxies a
fit to the sample of \citet{2003ApJ...591..764C} yields 286\,kpc. Any other
tracer population assumes values in between these extremes. 

Fig.\,\ref{fig:dispersion_models2} illustrates this scenario by displaying
the measurements from the centre to the cluster regime. We plot all
available velocity dispersion measurements: asterisks are the upper
envelope from Fig.\,\ref{fig:dispersion_models1}, filled circles represent
globular clusters from \citet{2011A&A...531A.119R} and
\citet{2011A&A...531A...4M}, filled squares represent the galaxy sample
from \citet{2003ApJ...591..764C}. We evaluated the velocity dispersions
of the galaxies employing the dispersion estimator from
\citet{1993ASPC...50.....D}.

One can see in Fig.\,\ref{fig:dispersion_models2} that beyond 40\,kpc the
upper envelope of the velocity dispersion of the stellar light smoothly
blends into the rising velocity dispersion profile of globular clusters,
reaching maximum values of $\sim$800\,km\,s$^{-1}$ at 100\,kpc. 
Outside this radius, the velocity dispersion profile of cluster galaxies 
stays flat with high values between 600 and 800\,km\,s$^{-1}$ out to
700\,kpc from the cluster centre. The upper panel in this figure shows
isotropic models, the lower panel the same radial
bias as in Fig.\,\ref{fig:dispersion_models1}. The solid (blue) lines are
models assuming as a mass profile the sum of the hydrodynamic X-ray
mass profile and the stellar mass in order to have a good approximation
for small radii, where the X-ray mass fails. The dashed (green) lines are
MOND models. The four values in parsecs to the left of the upper panel
correspond sequentially to the core radii. The red lines in both panels are
the cored Burkert halos from \citet{2011A&A...531A.119R} for the extreme
core radii. 

Even a mass profile like the one suggested by MOND may produce the
full range of observed velocity dispersions, depending on the properties
of the tracer population, which observationally are not directly accessible.
The cored dark halo of \citet{2011A&A...531A.119R} may owe its existence
only to the assumption of a uniform density profile of the tracer population. 

However, the cluster galaxy sample is not well described by the MOND
models. A thorough discussion is beyond the scope of our paper, but we
note that the classical assertion that galaxy clusters need dark matter
even with the validity of MOND may merit further investigation.
Galaxy clusters are anyway not in the deep MOND regime, but in the
transition region where some arbitrariness regarding the exact behaviour
exists (with the mass of Fig.\ref{fig:dispersion_models2}, the acceleration
at 100\,kpc is $1.3\times a_0$, at 1\,Mpc $0.2\times a_0$), with $a_0$
being the acceleration threshold in the MOND prescription 
($a_0\sim 10^{-8}$cm\,s$^{-2}$).

Furthermore, we probably oversimplify the MONDian model by assuming
a continuous mass distribution and neglecting galaxy interactions.
In addition, we assume that the X-ray gas constitutes the entire baryon
content. Finally, galaxy infall into a cluster does not start from zero velocity,
but already with velocities corresponding to the large scale structure,
that in turn must be studied under MOND conditions 
\citep{2016MNRAS.460.2571C}.

\section{Conclusions and summary}
\label{sec:conclusion}

NGC\,3311 is the central galaxy in the Hydra\,I galaxy cluster. Previous
work revealed much evidence for matter infall from the cluster environment
from galactocentric radii of about 5-10\,kpc outwards. As for other central
galaxies, a globally photometrically distinct halo is not discernible, but the
velocity dispersion of the stellar population, and at larger radii the velocity
dispersion of globular clusters, rise from the central value of about
180\,km\,s$^{-1}$ to about 800\,km\,s$^{-1}$ characterising the cluster's
galaxy population. In the framework of Jeans models this behaviour is
best explained by a large cored halo of dark matter. Seemingly conflicting
long-slit measurements of the velocity dispersion caused the wish for a
more complete measurements of the velocity field.      

In order to find kinematic signatures of sub-structures around NGC\,3311
we used FORS2 in MXU mode to mimic a coarse `IFU'.
Our novel approach of placing short slits in an onion shell-like pattern
around NGC\,3311 allowed us to measure its 2D large-scale kinematics
out to $\sim$4-5 effective radii ($\sim$38 kpc). This is the first time that
kinematic maps for a central galaxy are measured out to that radius.

These data show that the velocity field becomes inhomogeneous at
about 10\,kpc, slightly outside the effective radius of NGC\,3311.
The inner spheroid of NGC\,3311 can be considered as a slow rotator.
Only at larger radii the cumulative angular momentum is rising, however,
without showing an ordered rotation signal. The increased angular
momentum is probably due to the superposition of different kinematic
components that constitute the system inner spheroid-galaxy
halo-intracluster light (`cD halo') \citep{2018A&A...609A..78B}.
Also the velocity dispersion field loses its inner homogeneity and varies
as a function of radius and azimuthal angle, violating strict point symmetry.

Our results show that this complete spatial coverage is needed to avoid
biases of radial kinematic profiles from long-slit data and to draw
conclusions about the dynamical state of NGC\,3311 and the mass of
its dark matter halo. The conflict with earlier discrepant measurements 
finds its natural solution in the inhomogeneity of the kinematics as a
function of the azimuth.

Folding all measurements into a radial coordinate, the velocity dispersion
values lose their strict radial dependence at about 5-10\,kpc. Instead
they start to fill an area characterised by two well defined envelopes. The
lower envelope is a constant value of 200\,km\,s$^{-1}$, fitting well to the
MONDian expectation. The upper envelope can be written as,
$\sigma_{\rm upper}$[km\,s$^{-1}$]$=170\times(1+ (r/7.3)^{0.59})$ ($r$ in kpc),
which quite precisely connects the stars with the globular cluster population. 

Given the infall processes, we interpret this behaviour as the superposition
of various tracer populations. Tracer populations can probe strongly varying
potential differences within the cluster depending on their spatial distribution.
We explain the fact of a minimum and a maximum velocity dispersion at a
given radius by a minimally concentrated population (the cluster galaxies)
and a maximally concentrated population (the stars of NGC\,3311's inner
spheroid). Between these extremes, all dispersion values are possible.
In the transition zone we sample with increasing radius more and more
radially biased populations with a shallow distribution. A long-slit average
over all populations thus results in a rising velocity dispersion. 

Under these circumstances, the potential cannot be precisely constrained
except for the very inner region (where baryons dominate) and for the large
scale galaxy potential. Previous statements about dark halos are probably 
not valid.
We show with Jeans models that a range of mass profiles 
can generate all observed projected velocity dispersions depending on the
core radius of their distribution. The MOND mass profile is also capable of
producing the inner dispersion values, but fails in case of the cluster wide
potential. We caution, however, that a more trustworthy test needs to be
done in order to probe the formation of the galaxy cluster under MOND
conditions.

With the increasing availability of large field IFUs more galaxies will be
observed. We predict that the increasing scatter of dispersion values will
emerge as a characteristic
for those central galaxies with rising velocity dispersion profiles.
The cored dark halo of \citet{2011A&A...531A.119R}
may owe its existence only to the assumption of a uniform density profile
of the tracer population. 

\begin{acknowledgements}
We would like to thank the anonymous referee for his/her very valuable
comments.
Based on observations collected at the European Organisation for Astronomical 
Research in the Southern Hemisphere under ESO programme 088.B-0448(B).
This research has made
use of the NASA/IPAC Extragalactic Database (NED) which is operated by the
Jet Propulsion Laboratory, California Institute of Technology, under contract
with the National Aeronautics and Space Administration. 
T.R. acknowledges support from the BASAL Centro de Astrof\'{\i}sica y
Tecnologias Afines (CATA) projects PFB-06/2007 and AFB-170002, and from
FONDECYT project Nr. 1100620. T.R. thanks ESO/Garching for a science
visitorship during May-September 2016.
C.E.B and C.M.dO. thank the S\~{a}o Paulo Research Foundation (FAPESP)
funding (grants 2011/21325-0, 2011/51680-6, 2012/22676-3 and 2016/12331-0).
\end{acknowledgements}

\bibliographystyle{aa} 
\bibliography{biblio_n3311}

\begin{appendix}
\section{Results table}
\onecolumn

\begin{longtab}
\begin{landscape}
\small
\begin{longtable}{lllrrlrrrrccc}
\caption{\label{tab:kinematics} Measured kinematic properties around NGC\,3311.}\\
\hline\hline
ID & RA & DEC & R & PA & $V_{\rm{LOS}}$ & $\sigma_{\rm{LOS}}$ & $h_3$ & $h_4$ & S/N & adeg & mdeg & flag \\
 & (J2000) & (J2000) & (kpc) & ($^{\rm o}$) & (km\,s$^{-1}$) & (km\,s$^{-1}$) & & & (\AA$^{-1}$) & & & \\
(1) & (2) & (3) & (4) & (5) & (6) & (7) & (8) & (9) & (10) & (11) & (12) & (13)\\
\hline
\endfirsthead
\caption{continued.}\\
\hline\hline
ID & RA & DEC & R & PA & $V_{\rm{LOS}}$ & $\sigma_{\rm{LOS}}$ & $h_3$ & $h_4$ & S/N & adeg & mdeg & flag \\
 & (J2000) & (J2000) & (arcsec) & ($^{\rm o}$) & (km\,s$^{-1}$) & (km\,s$^{-1}$) &  & & (\AA$^{-1}$) & & & \\
(1) & (2) & (3) & (4) & (5) & (6) & (7) & (8) & (9) & (10) & (11) & (12) & (13)\\
\hline
\endhead
\hline
\endfoot
cen1\_s14a & 159h10m08.4s & -27d33m34.2s & 28.8 & 130.4 & $4895\pm7$ & $54\pm13$ & $-0.132\pm0.110$ & $0.035\pm0.156$ & 92.5 & 6 & 4 & 3 \\
cen1\_s14b & 159h10m04.8s & -27d33m36.4s & 29.6 & 130.9 & $4858\pm6$ & $69\pm8$ & $0.008\pm0.072$ & $-0.047\pm0.068$ & 54.7 & 6 & 4 & 3 \\
cen1\_s14c & 159h10m12.0s & -27d33m32.0s & 28.0 & 129.8 & $4867\pm6$ & $70\pm10$ & $-0.103\pm0.066$ & $0.001\pm0.100$ & 55.2 & 6 & 4 & 3 \\
cen1\_s18 & 159h09m43.2s & -27d32m33.4s & 19.2 & 111.3 & $4136\pm42$ & $322\pm47$ & $-0.056\pm0.053$ & $-0.008\pm0.199$ & 14.3 & 6 & 4 & 1 \\
cen1\_s19 & 159h09m18.0s & -27d32m04.6s & 21.4 & 99.7 & $4031\pm26$ & $214\pm30$ & $-0.064\pm0.052$ & $-0.002\pm0.113$ & 13.4 & 6 & 4 & 2 \\
cen1\_s20 & 159h10m01.2s & -27d32m24.7s & 14.5 & 108.0 & $3970\pm24$ & $281\pm28$ & $0.065\pm0.061$ & $0.101\pm0.056$ & 19.1 & 6 & 4 & 1 \\
cen1\_s21 & 159h09m57.6s & -27d32m10.7s & 13.0 & 102.3 & $3796\pm22$ & $348\pm20$ & $-0.175\pm0.035$ & $0.013\pm0.048$ & 24.0 & 6 & 4 & 1 \\
cen1\_s23 & 159h10m01.2s & -27d31m56.6s & 10.6 & 96.3 & $3866\pm9$ & $280\pm12$ & $-0.067\pm0.025$ & $0.138\pm0.026$ & 46.9 & 6 & 4 & 1 \\
cen1\_s24 & 159h10m15.6s & -27d31m52.3s & 6.9 & 94.5 & $3874\pm9$ & $306\pm11$ & $-0.045\pm0.022$ & $0.087\pm0.024$ & 44.6 & 6 & 4 & 1 \\
cen1\_s25 & 159h10m01.2s & -27d31m34.3s & 10.2 & 86.7 & $3895\pm9$ & $262\pm10$ & $-0.142\pm0.024$ & $0.015\pm0.029$ & 52.1 & 6 & 4 & 1 \\
cen1\_s26 & 159h10m15.6s & -27d31m35.8s & 6.6 & 87.3 & $3884\pm6$ & $287\pm7$ & $0.073\pm0.015$ & $0.122\pm0.017$ & 73.7 & 6 & 4 & 1 \\
cen1\_s27 & 159h10m33.6s & -27d31m42.2s & 2.0 & 90.1 & $3851\pm2$ & $203\pm3$ & $0.039\pm0.009$ & $0.085\pm0.010$ & 131.1 & 6 & 4 & 1 \\
cen1\_s28 & 159h10m08.4s & -27d31m16.7s & 10.3 & 79.1 & $3886\pm16$ & $335\pm16$ & $-0.024\pm0.032$ & $0.017\pm0.037$ & 41.2 & 6 & 4 & 1 \\
cen1\_s29 & 159h10m48.0s & -27d31m34.7s & 2.3 & 86.8 & $3879\pm2$ & $195\pm3$ & $-0.014\pm0.009$ & $0.062\pm0.010$ & 120.6 & 6 & 4 & 1 \\
cen1\_s29a & 159h10m46.9s & -27d31m35.6s & 2.0 & 87.2 & $3879\pm2$ & $195\pm3$ & $-0.012\pm0.009$ & $0.065\pm0.010$ & 121.1 & 6 & 4 & 1 \\
cen1\_s29b & 159h10m49.1s & -27d31m33.7s & 2.7 & 86.4 & $3885\pm2$ & $193\pm3$ & $-0.023\pm0.009$ & $0.098\pm0.011$ & 113.3 & 6 & 4 & 1 \\
cen1\_s30 & 159h10m26.4s & -27d31m09.5s & 8.9 & 76.2 & $3914\pm11$ & $349\pm14$ & $-0.019\pm0.023$ & $0.102\pm0.024$ & 62.3 & 6 & 4 & 1 \\
cen1\_s31 & 159h10m48.0s & -27d31m17.4s & 6.2 & 79.4 & $3926\pm5$ & $253\pm6$ & $0.002\pm0.017$ & $0.028\pm0.017$ & 63.2 & 6 & 4 & 1 \\
cen1\_s32 & 159h10m44.4s & -27d31m05.9s & 8.9 & 74.7 & $3943\pm13$ & $459\pm15$ & $0.053\pm0.025$ & $0.268\pm0.024$ & 54.0 & 6 & 4 & 1 \\
cen1\_s33 & 159h11m02.4s & -27d31m10.6s & 9.2 & 76.6 & $3955\pm9$ & $307\pm12$ & $-0.039\pm0.023$ & $0.120\pm0.024$ & 63.5 & 6 & 4 & 1 \\
cen1\_s34 & 159h11m02.4s & -27d30m59.4s & 11.6 & 72.1 & $3935\pm17$ & $402\pm19$ & $0.012\pm0.029$ & $0.050\pm0.028$ & 40.0 & 6 & 4 & 1 \\
cen1\_s35 & 159h10m58.8s & -27d30m47.5s & 14.0 & 67.6 & $3888\pm14$ & $283\pm17$ & $-0.013\pm0.039$ & $0.078\pm0.036$ & 33.4 & 6 & 4 & 1 \\
cen1\_s36 & 159h10m48.0s & -27d30m27.4s & 18.4 & 60.5 & $3906\pm25$ & $250\pm28$ & $0.010\pm0.078$ & $0.027\pm0.058$ & 24.1 & 8 & 2 & 1 \\
cen1\_s37 & 159h11m31.2s & -27d30m49.7s & 17.7 & 68.4 & $3948\pm30$ & $411\pm29$ & $-0.028\pm0.043$ & $-0.043\pm0.051$ & 21.7 & 6 & 4 & 1 \\
cen2\_s21 & 159h11m45.6s & -27d31m01.9s & 18.5 & 73.1 & $3878\pm33$ & $452\pm37$ & $-0.051\pm0.045$ & $0.043\pm0.046$ & 25.5 & 6 & 4 & 1 \\
cen2\_s22 & 159h11m45.6s & -27d31m12.7s & 17.2 & 77.5 & $3860\pm31$ & $443\pm40$ & $-0.023\pm0.052$ & $0.162\pm0.043$ & 24.0 & 6 & 4 & 1 \\
cen2\_s23 & 159h11m31.2s & -27d31m12.0s & 14.2 & 77.2 & $3896\pm18$ & $395\pm22$ & $-0.003\pm0.033$ & $0.109\pm0.031$ & 35.4 & 6 & 4 & 1 \\
cen2\_s25 & 159h11m24.0s & -27d31m24.2s & 11.2 & 82.3 & $3965\pm10$ & $378\pm12$ & $0.006\pm0.025$ & $0.285\pm0.025$ & 42.3 & 6 & 4 & 1 \\
cen2\_s26 & 159h11m24.0s & -27d31m38.6s & 10.4 & 88.5 & $3899\pm14$ & $305\pm15$ & $-0.027\pm0.032$ & $0.035\pm0.033$ & 32.8 & 6 & 4 & 1 \\
cen2\_s27 & 159h11m09.6s & -27d31m36.5s & 6.9 & 87.6 & $3917\pm7$ & $268\pm8$ & $0.015\pm0.019$ & $0.086\pm0.019$ & 67.8 & 6 & 4 & 1 \\
cen2\_s28 & 159h11m20.4s & -27d31m52.0s & 9.8 & 94.3 & $3859\pm14$ & $380\pm17$ & $-0.120\pm0.026$ & $0.143\pm0.029$ & 42.8 & 6 & 4 & 1 \\
cen2\_s29 & 159h11m06.0s & -27d31m48.4s & 6.1 & 92.8 & $3872\pm6$ & $295\pm8$ & $-0.002\pm0.016$ & $0.139\pm0.017$ & 68.8 & 6 & 4 & 1 \\
cen2\_s30 & 159h10m48.0s & -27d31m41.9s & 1.5 & 90.0 & $3862\pm2$ & $215\pm3$ & $-0.007\pm0.008$ & $0.082\pm0.009$ & 109.8 & 6 & 4 & 1 \\
cen2\_s31 & 159h11m13.2s & -27d32m06.4s & 9.8 & 100.5 & $3942\pm10$ & $280\pm13$ & $0.010\pm0.029$ & $0.124\pm0.029$ & 24.4 & 6 & 4 & 1 \\
cen2\_s32 & 159h10m37.2s & -27d31m47.3s & 1.7 & 92.3 & $3848\pm2$ & $201\pm3$ & $-0.006\pm0.009$ & $0.095\pm0.011$ & 114.7 & 6 & 4 & 1 \\
cen2\_s32a & 159h10m36.2s & -27d31m48.1s & 2.0 & 92.6 & $3851\pm2$ & $185\pm3$ & $-0.020\pm0.009$ & $0.094\pm0.011$ & 113.8 & 6 & 4 & 1 \\
cen2\_s32b & 159h10m38.2s & -27d31m46.5s & 1.4 & 91.9 & $3848\pm2$ & $200\pm3$ & $-0.005\pm0.010$ & $0.097\pm0.011$ & 115.5 & 6 & 4 & 1 \\
cen2\_s33 & 159h11m02.4s & -27d32m13.2s & 9.2 & 103.3 & $3933\pm10$ & $276\pm11$ & $-0.029\pm0.027$ & $0.032\pm0.028$ & 37.6 & 6 & 4 & 1 \\
cen2\_s34 & 159h10m44.4s & -27d32m08.2s & 6.5 & 101.2 & $3867\pm5$ & $224\pm7$ & $0.045\pm0.019$ & $0.067\pm0.021$ & 66.1 & 6 & 4 & 1 \\
cen2\_s35 & 159h10m44.4s & -27d32m18.2s & 8.9 & 105.4 & $3874\pm10$ & $270\pm13$ & $0.038\pm0.030$ & $0.126\pm0.030$ & 37.7 & 6 & 4 & 1 \\
cen2\_s36 & 159h10m26.4s & -27d32m13.2s & 8.6 & 103.3 & $3895\pm9$ & $278\pm12$ & $0.078\pm0.026$ & $0.114\pm0.028$ & 48.5 & 6 & 4 & 1 \\
cen2\_s37 & 159h10m30.0s & -27d32m24.7s & 10.9 & 108.0 & $3895\pm15$ & $273\pm15$ & $-0.179\pm0.034$ & $0.073\pm0.043$ & 27.8 & 6 & 4 & 1 \\
cen2\_s38 & 159h10m15.6s & -27d32m23.6s & 12.1 & 107.5 & $3891\pm14$ & $241\pm18$ & $-0.074\pm0.045$ & $0.110\pm0.044$ & 30.4 & 6 & 4 & 1 \\
cen2\_s39 & 159h10m22.8s & -27d32m41.3s & 15.3 & 114.2 & $3926\pm56$ & $558\pm74$ & $0.052\pm0.075$ & $0.133\pm0.054$ & 16.6 & 6 & 4 & 1 \\
cen2\_s40 & 159h09m57.6s & -27d32m32.6s & 16.5 & 111.0 & $3827\pm46$ & $426\pm53$ & $-0.061\pm0.063$ & $0.073\pm0.069$ & 16.0 & 8 & 2 & 1 \\
cen2\_s45a & 159h10m26.4s & -27d33m45.0s & 30.5 & 133.0 & $4531\pm3$ & $59\pm5$ & $0.058\pm0.037$ & $0.090\pm0.059$ & 86.5 & 6 & 4 & 3 \\
cen2\_s45b & 159h10m22.8s & -27d33m47.5s & 31.2 & 133.6 & $4564\pm4$ & $69\pm8$ & $0.103\pm0.046$ & $0.044\pm0.081$ & 39.4 & 6 & 4 & 3 \\
cen2\_s45c & 159h10m30.0s & -27d33m42.5s & 29.8 & 132.4 & $4566\pm7$ & $70\pm10$ & $0.042\pm0.077$ & $0.093\pm0.084$ & 44.4 & 6 & 4 & 3 \\
inn1\_s15 & 159h08m45.6s & -27d32m26.2s & 30.6 & 108.5 & $3839\pm154$ & $203\pm432$ & -- & -- & 12.5 & 8 & 2 & 1 \\
inn1\_s16 & 159h09m25.2s & -27d32m43.1s & 24.1 & 114.9 & $4131\pm65$ & $417\pm70$ & $0.015\pm0.103$ & $0.033\pm0.055$ & 16.8 & 6 & 4 & 1 \\
inn1\_s17 & 159h09m39.6s & -27d32m40.6s & 21.0 & 113.9 & $4092\pm47$ & $403\pm52$ & $-0.108\pm0.059$ & $0.053\pm0.078$ & 15.0 & 6 & 4 & 1 \\
inn1\_s18 & 159h09m21.6s & -27d32m15.7s & 21.4 & 104.3 & $3902\pm44$ & $420\pm30$ & $-0.109\pm0.069$ & $-0.153\pm0.060$ & 14.2 & 6 & 4 & 1 \\
inn1\_s19 & 159h09m39.6s & -27d32m19.3s & 17.8 & 105.8 & $3911\pm32$ & $312\pm43$ & $-0.020\pm0.080$ & $0.152\pm0.053$ & 18.7 & 6 & 4 & 1 \\
inn1\_s20 & 159h09m14.4s & -27d31m51.2s & 21.6 & 94.0 & $4117\pm18$ & $235\pm23$ & $0.040\pm0.061$ & $0.077\pm0.053$ & 26.4 & 6 & 4 & 2 \\
inn1\_s21 & 159h09m36.0s & -27d31m57.0s & 16.6 & 96.5 & $3908\pm23$ & $320\pm27$ & $-0.102\pm0.049$ & $0.096\pm0.051$ & 29.1 & 6 & 4 & 1 \\
inn1\_s23 & 159h09m36.0s & -27d31m37.6s & 16.2 & 88.1 & $4011\pm18$ & $336\pm23$ & $-0.012\pm0.043$ & $0.219\pm0.037$ & 25.8 & 6 & 4 & 1 \\
inn1\_s24 & 159h09m39.6s & -27d31m26.4s & 15.8 & 83.3 & $4039\pm34$ & $360\pm45$ & $0.009\pm0.074$ & $0.153\pm0.050$ & 18.2 & 6 & 4 & 1 \\
inn1\_s25 & 159h09m28.8s & -27d31m10.2s & 19.6 & 76.4 & $4062\pm7$ & $210\pm8$ & $-0.004\pm0.026$ & $0.043\pm0.026$ & 62.2 & 6 & 4 & 2 \\
inn1\_s26 & 159h09m46.8s & -27d31m12.4s & 15.4 & 77.3 & $3987\pm15$ & $320\pm20$ & $-0.034\pm0.036$ & $0.180\pm0.035$ & 29.1 & 6 & 4 & 1 \\
inn1\_s27 & 159h09m32.4s & -27d30m49.7s & 21.4 & 68.4 & $4061\pm10$ & $225\pm12$ & $-0.059\pm0.033$ & $0.004\pm0.040$ & 43.1 & 6 & 4 & 2 \\
inn1\_s28 & 159h10m01.2s & -27d30m58.0s & 14.7 & 71.5 & $3940\pm13$ & $241\pm16$ & $-0.087\pm0.037$ & $0.052\pm0.045$ & 29.2 & 6 & 4 & 1 \\
inn1\_s29 & 159h09m50.4s & -27d30m39.2s & 20.0 & 64.6 & $3978\pm24$ & $276\pm20$ & $-0.047\pm0.057$ & $-0.069\pm0.061$ & 18.1 & 6 & 4 & 2 \\
inn1\_s30 & 159h10m19.2s & -27d30m47.9s & 14.4 & 67.7 & $3907\pm21$ & $388\pm24$ & $0.017\pm0.038$ & $0.082\pm0.034$ & 29.0 & 6 & 4 & 1 \\
inn1\_s31 & 159h10m08.4s & -27d30m29.5s & 19.6 & 61.2 & $4070\pm41$ & $429\pm47$ & $0.120\pm0.059$ & $0.111\pm0.067$ & 21.1 & 6 & 4 & 1 \\
inn1\_s32 & 159h10m44.4s & -27d30m44.6s & 14.1 & 66.5 & $3894\pm17$ & $351\pm22$ & $0.047\pm0.036$ & $0.130\pm0.034$ & 43.9 & 6 & 4 & 1 \\
inn1\_s33 & 159h10m33.6s & -27d30m23.8s & 19.3 & 59.3 & $4021\pm32$ & $416\pm43$ & $0.048\pm0.055$ & $0.114\pm0.048$ & 25.0 & 6 & 4 & 1 \\
inn1\_s34 & 159h11m16.8s & -27d30m45.4s & 16.4 & 66.8 & $3961\pm23$ & $356\pm22$ & $0.010\pm0.043$ & $-0.007\pm0.044$ & 30.5 & 6 & 4 & 1 \\
inn1\_s35 & 159h11m09.6s & -27d30m29.9s & 19.0 & 61.3 & $3975\pm25$ & $284\pm25$ & $-0.088\pm0.046$ & $-0.012\pm0.072$ & 33.7 & 6 & 4 & 1 \\
inn1\_s36 & 159h11m38.4s & -27d30m37.8s & 21.0 & 64.1 & $4015\pm29$ & $329\pm33$ & $-0.039\pm0.061$ & $0.047\pm0.056$ & 25.2 & 6 & 4 & 1 \\
inn1\_s37 & 159h11m38.4s & -27d30m23.8s & 23.7 & 59.3 & $3993\pm34$ & $312\pm40$ & $0.004\pm0.085$ & $0.045\pm0.051$ & 20.3 & 6 & 4 & 1 \\
inn1\_s39 & 159h12m00.0s & -27d30m08.6s & 29.9 & 54.7 & $4144\pm35$ & $250\pm44$ & $0.021\pm0.114$ & $0.159\pm0.055$ & 19.9 & 6 & 4 & 1 \\
inn2\_s18 & 159h11m52.8s & -27d30m49.7s & 21.7 & 68.4 & $3927\pm32$ & $413\pm32$ & $0.024\pm0.042$ & $0.010\pm0.054$ & 22.6 & 6 & 4 & 1 \\
inn2\_s19 & 159h12m00.0s & -27d31m05.5s & 21.2 & 74.5 & $3711\pm39$ & $478\pm35$ & $-0.009\pm0.057$ & $-0.046\pm0.043$ & 17.7 & 8 & 2 & 1 \\
inn2\_s20 & 159h11m38.4s & -27d31m01.6s & 17.1 & 73.0 & $3820\pm36$ & $603\pm43$ & $0.077\pm0.040$ & $0.078\pm0.042$ & 30.3 & 6 & 4 & 1 \\
inn2\_s21 & 159h12m07.2s & -27d31m34.0s & 21.1 & 86.5 & $4256\pm84$ & $441\pm98$ & $0.009\pm0.136$ & $0.071\pm0.058$ & 17.5 & 6 & 4 & 1 \\
inn2\_s22 & 159h11m45.6s & -27d31m28.6s & 16.0 & 84.2 & $3851\pm19$ & $220\pm21$ & $-0.055\pm0.048$ & $0.029\pm0.073$ & 20.8 & 6 & 4 & 1 \\
inn2\_s23 & 159h12m07.2s & -27d31m55.2s & 21.2 & 95.7 & $3908\pm33$ & $318\pm28$ & $-0.175\pm0.051$ & $-0.038\pm0.078$ & 13.3 & 6 & 4 & 1 \\
inn2\_s24 & 159h11m45.6s & -27d31m53.4s & 15.9 & 94.9 & $3921\pm46$ & $584\pm60$ & $-0.043\pm0.060$ & $0.140\pm0.050$ & 18.6 & 8 & 2 & 1 \\
inn2\_s25 & 159h12m00.0s & -27d32m13.9s & 20.7 & 103.6 & $3842\pm47$ & $321\pm43$ & $-0.022\pm0.115$ & $-0.051\pm0.062$ & 14.8 & 6 & 4 & 1 \\
inn2\_s26 & 159h11m38.4s & -27d32m09.6s & 15.5 & 101.8 & $3955\pm36$ & $413\pm39$ & $-0.142\pm0.047$ & $0.072\pm0.062$ & 20.7 & 6 & 4 & 1 \\
inn2\_s27 & 159h11m45.6s & -27d32m30.1s & 19.6 & 110.0 & $3292\pm104$ & -- & $-0.164\pm0.059$ & $0.105\pm0.080$ & 13.8 & 6 & 4 & 1 \\
inn2\_s28 & 159h11m20.4s & -27d32m25.8s & 14.3 & 108.4 & $3972\pm24$ & $426\pm32$ & $0.039\pm0.040$ & $0.113\pm0.036$ & 27.6 & 6 & 4 & 1 \\
inn2\_s29 & 159h11m31.2s & -27d32m44.2s & 19.5 & 115.2 & $3755\pm32$ & $204\pm33$ & $0.040\pm0.088$ & $-0.027\pm0.116$ & 16.6 & 6 & 4 & 1 \\
inn2\_s30 & 159h11m02.4s & -27d32m35.5s & 14.1 & 112.1 & $3800\pm27$ & $404\pm24$ & $-0.038\pm0.040$ & $-0.038\pm0.046$ & 27.3 & 6 & 4 & 1 \\
inn2\_s31 & 159h11m09.6s & -27d32m51.7s & 18.4 & 117.9 & $3859\pm60$ & $480\pm62$ & $-0.050\pm0.051$ & $0.002\pm0.089$ & 14.6 & 6 & 4 & 1 \\
inn2\_s32 & 159h10m37.2s & -27d32m37.0s & 13.6 & 112.6 & $3866\pm27$ & $427\pm30$ & $0.066\pm0.039$ & $0.050\pm0.044$ & 24.7 & 6 & 4 & 1 \\
inn2\_s33 & 159h10m44.4s & -27d32m55.3s & 18.0 & 119.1 & $4028\pm48$ & $440\pm56$ & $-0.086\pm0.073$ & $0.135\pm0.062$ & 14.9 & 6 & 4 & 1 \\
inn2\_s34 & 159h10m08.4s & -27d32m39.5s & 16.4 & 113.6 & $3918\pm34$ & $306\pm45$ & $-0.016\pm0.084$ & $0.155\pm0.052$ & 20.0 & 6 & 4 & 1 \\
inn2\_s35 & 159h10m19.2s & -27d32m55.7s & 19.0 & 119.2 & $3734\pm41$ & $280\pm40$ & $-0.040\pm0.097$ & $-0.030\pm0.091$ & 14.4 & 6 & 4 & 1 \\
inn2\_s36 & 159h09m54.0s & -27d32m46.7s & 19.8 & 116.1 & $3982\pm23$ & $274\pm27$ & $-0.081\pm0.067$ & $0.262\pm0.051$ & 18.7 & 6 & 4 & 1 \\
inn2\_s37 & 159h10m30.0s & -27d33m26.3s & 25.8 & 128.3 & $4488\pm99$ & -- & $0.019\pm0.049$ & $0.025\pm0.119$ & 16.7 & 6 & 4 & 1 \\
inn2\_s38 & 159h09m57.6s & -27d33m08.3s & 23.8 & 123.2 & $3445\pm24$ & -- & $0.034\pm0.106$ & -- & 21.4 & 6 & 4 & 1 \\
inn2\_s39a & 159h10m19.2s & -27d33m41.0s & 29.8 & 132.1 & $4747\pm5$ & $114\pm7$ & $0.010\pm0.034$ & $0.053\pm0.044$ & 146.1 & 6 & 4 & 3 \\
inn2\_s39b & 159h10m15.6s & -27d33m42.1s & 30.2 & 132.3 & $4749\pm4$ & $91\pm8$ & $0.012\pm0.041$ & $0.080\pm0.064$ & 175.8 & 6 & 4 & 3 \\
inn2\_s39c & 159h10m19.2s & -27d33m39.6s & 29.4 & 131.7 & $4759\pm5$ & $88\pm10$ & $-0.008\pm0.044$ & $0.097\pm0.077$ & 171.1 & 6 & 4 & 3 \\
inn2\_s39d & 159h10m15.6s & -27d33m44.3s & 30.7 & 132.8 & $4749\pm3$ & $71\pm5$ & $0.022\pm0.030$ & $0.084\pm0.047$ & 110.5 & 6 & 4 & 3 \\
inn2\_s39e & 159h10m22.8s & -27d33m36.7s & 28.6 & 131.0 & $4762\pm3$ & $81\pm4$ & $-0.021\pm0.026$ & $0.057\pm0.038$ & 100.6 & 6 & 4 & 3 \\
out1\_s13 & 159h08m56.4s & -27d33m15.5s & 34.6 & 125.3 & $4113\pm117$ & $706\pm87$ & $-0.043\pm0.133$ & $-0.125\pm0.056$ & 10.0 & 6 & 4 & 1 \\
out1\_s14 & 159h08m42.0s & -27d32m51.7s & 34.1 & 117.9 & $3745\pm36$ & $336\pm38$ & $-0.017\pm0.075$ & $0.051\pm0.053$ & 12.5 & 6 & 4 & 1 \\
out1\_s15 & 159h08m20.4s & -27d32m22.2s & 36.2 & 107.0 & $4002\pm78$ & $358\pm61$ & $-0.098\pm0.106$ & $-0.057\pm0.130$ & 15.5 & 6 & 4 & 1 \\
out1\_s16 & 159h08m49.2s & -27d32m30.8s & 30.2 & 110.3 & $4009\pm50$ & $376\pm56$ & $-0.161\pm0.076$ & $0.116\pm0.091$ & 11.8 & 6 & 4 & 1 \\
out1\_s17 & 159h08m13.2s & -27d31m49.1s & 36.6 & 93.1 & $4211\pm20$ & $290\pm23$ & $0.193\pm0.046$ & $0.108\pm0.058$ & 18.1 & 6 & 4 & 2 \\
out1\_s18 & 159h08m38.4s & -27d31m55.6s & 30.5 & 95.9 & $4093\pm12$ & $214\pm16$ & $-0.056\pm0.045$ & $0.158\pm0.043$ & 38.7 & 6 & 4 & 2 \\
out1\_s19 & 159h08m13.2s & -27d31m23.2s & 36.8 & 81.9 & $4096\pm11$ & $235\pm12$ & $-0.170\pm0.032$ & $0.075\pm0.040$ & 42.9 & 6 & 4 & 2 \\
out1\_s20 & 159h08m38.4s & -27d31m28.6s & 30.5 & 84.2 & $4097\pm3$ & $196\pm5$ & $0.056\pm0.014$ & $0.071\pm0.016$ & 77.4 & 6 & 4 & 2 \\
out1\_s21 & 159h08m16.8s & -27d30m58.7s & 37.2 & 71.8 & $4095\pm6$ & $180\pm8$ & $-0.036\pm0.026$ & $0.086\pm0.029$ & 46.2 & 6 & 4 & 2 \\
out1\_s22 & 159h08m20.4s & -27d30m47.5s & 37.3 & 67.6 & $4056\pm9$ & $240\pm12$ & $-0.102\pm0.029$ & $0.098\pm0.032$ & 36.2 & 6 & 4 & 2 \\
out1\_s23 & 159h08m45.6s & -27d30m36.4s & 32.8 & 63.5 & $4056\pm5$ & $212\pm7$ & $-0.025\pm0.019$ & $0.074\pm0.021$ & 47.4 & 6 & 4 & 2 \\
out1\_s24 & 159h08m45.6s & -27d30m15.5s & 35.6 & 56.7 & $4071\pm18$ & $283\pm21$ & $0.137\pm0.045$ & $0.160\pm0.048$ & 18.3 & 6 & 4 & 2 \\
out1\_s25 & 159h09m21.6s & -27d30m23.8s & 27.6 & 59.3 & $4084\pm13$ & $225\pm16$ & $0.115\pm0.040$ & $0.119\pm0.045$ & 28.1 & 6 & 4 & 2 \\
out1\_s26 & 159h09m07.2s & -27d29m58.9s & 34.4 & 52.0 & $4141\pm30$ & $218\pm37$ & $0.105\pm0.087$ & $0.091\pm0.100$ & 13.4 & 6 & 4 & 2 \\
out1\_s27 & 159h09m14.4s & -27d29m52.1s & 34.5 & 50.2 & $5280\pm95$ & -- & $0.127\pm0.050$ & $0.032\pm0.055$ & 10.0 & 6 & 4 & 2 \\
out1\_s28 & 159h09m46.8s & -27d30m02.2s & 28.0 & 52.9 & $4052\pm38$ & $336\pm49$ & $-0.041\pm0.085$ & $0.161\pm0.052$ & 9.6 & 6 & 4 & 1 \\
out1\_s29 & 159h09m36.0s & -27d29m40.6s & 34.0 & 47.4 & $4940\pm112$ & $526\pm96$ & $0.039\pm0.152$ & $-0.094\pm0.085$ & 6.8 & 8 & 2 & 1 \\
out1\_s30 & 159h10m15.6s & -27d29m56.8s & 26.7 & 51.4 & $4296\pm85$ & $635\pm77$ & $0.040\pm0.077$ & $-0.075\pm0.080$ & 13.8 & 6 & 4 & 1 \\
out1\_s31 & 159h10m01.2s & -27d29m32.6s & 33.3 & 45.6 & $3828\pm66$ & $342\pm47$ & $0.068\pm0.141$ & $-0.154\pm0.081$ & 7.4 & 6 & 4 & 1 \\
out1\_s32 & 159h10m48.0s & -27d29m53.9s & 26.6 & 50.7 & $3974\pm28$ & $275\pm31$ & $-0.013\pm0.077$ & $0.047\pm0.055$ & 21.7 & 6 & 4 & 1 \\
out1\_s33 & 159h10m33.6s & -27d29m28.7s & 32.8 & 44.7 & $4485\pm107$ & $533\pm82$ & $0.056\pm0.125$ & $-0.120\pm0.091$ & 8.9 & 8 & 2 & 1 \\
out1\_s34 & 159h11m34.8s & -27d30m01.1s & 28.0 & 52.6 & $3668\pm58$ & $846\pm67$ & $-0.180\pm0.050$ & $0.155\pm0.053$ & 20.9 & 6 & 4 & 1 \\
out1\_s35 & 159h11m16.8s & -27d29m32.3s & 33.0 & 45.5 & $3613\pm50$ & $292\pm38$ & $-0.007\pm0.130$ & $-0.107\pm0.056$ & 9.7 & 6 & 4 & 1 \\
out1\_s36 & 159h12m03.6s & -27d29m53.9s & 33.3 & 50.7 & $3844\pm51$ & $348\pm53$ & $-0.055\pm0.078$ & $0.009\pm0.093$ & 13.7 & 6 & 4 & 1 \\
out2\_s13 & 159h12m25.2s & -27d30m10.8s & 33.9 & 55.3 & $3878\pm108$ & $1000\pm119$ & $-0.193\pm0.063$ & $0.151\pm0.081$ & 17.7 & 6 & 4 & 1 \\
out2\_s14 & 159h13m04.8s & -27d30m54.4s & 37.0 & 70.1 & $3546\pm57$ & $451\pm48$ & $-0.004\pm0.085$ & $-0.070\pm0.051$ & 11.0 & 6 & 4 & 1 \\
out2\_s15 & 159h12m21.6s & -27d30m36.4s & 29.3 & 63.5 & $3977\pm32$ & $437\pm40$ & $-0.079\pm0.048$ & $0.124\pm0.046$ & 20.6 & 6 & 4 & 1 \\
out2\_s16 & 159h13m12.0s & -27d31m26.8s & 37.1 & 83.4 & $3532\pm48$ & $403\pm37$ & $-0.096\pm0.074$ & $-0.088\pm0.068$ & 11.3 & 6 & 4 & 1 \\
out2\_s17 & 159h12m43.2s & -27d31m18.8s & 30.4 & 80.0 & $3883\pm25$ & $116\pm37$ & $0.065\pm0.069$ & $-0.002\pm0.250$ & 15.5 & 6 & 4 & 1 \\
out2\_s18 & 159h13m12.0s & -27d31m53.4s & 37.0 & 94.9 & $4904\pm79$ & -- & $0.156\pm0.055$ & $0.148\pm0.060$ & 12.0 & 6 & 4 & 1 \\
out2\_s19 & 159h12m43.2s & -27d31m47.3s & 29.8 & 92.3 & $3678\pm105$ & -- & $-0.134\pm0.053$ & $0.025\pm0.109$ & 14.5 & 6 & 4 & 1 \\
out2\_s20 & 159h13m01.2s & -27d32m08.9s & 34.9 & 101.5 & $3948\pm70$ & $630\pm73$ & $0.094\pm0.051$ & $-0.007\pm0.086$ & 15.2 & 8 & 2 & 1 \\
out2\_s21 & 159h12m39.6s & -27d32m06.0s & 29.5 & 100.3 & $3434\pm67$ & $680\pm73$ & $-0.012\pm0.066$ & $0.051\pm0.048$ & 14.5 & 6 & 4 & 1 \\
out2\_s22 & 159h13m01.2s & -27d32m34.4s & 36.6 & 111.7 & $3591\pm78$ & -- & $0.036\pm0.046$ & $-0.026\pm0.080$ & 5.7 & 6 & 4 & 1 \\
out2\_s23 & 159h12m54.0s & -27d32m46.0s & 36.1 & 115.9 & $3598\pm139$ & -- & $0.149\pm0.086$ & $-0.12\pm0.116$ & 4.6 & 6 & 4 & 1 \\
out2\_s24 & 159h12m28.8s & -27d32m37.3s & 29.6 & 112.8 & $2261\pm93$ & -- & $0.085\pm0.051$ & $0.106\pm0.051$ & 6.6 & 6 & 4 & 1 \\
out2\_s25 & 159h12m39.6s & -27d33m06.5s & 35.6 & 122.6 & $3923\pm300$ & -- & -- & -- & 7.8 & 6 & 4 & 1 \\
out2\_s26 & 159h12m03.6s & -27d32m59.3s & 27.6 & 120.4 & $3795\pm64$ & -- & $0.040\pm0.041$ & $0.023\pm0.063$ & 12.8 & 8 & 2 & 1 \\
out2\_s27 & 159h12m18.0s & -27d33m24.5s & 34.5 & 127.8 & $4034\pm300$ & -- & -- & -- & 5.2 & 6 & 4 & 1 \\
out2\_s28 & 159h11m45.6s & -27d33m12.2s & 27.2 & 124.4 & $3765\pm94$ & $535\pm80$ & $0.014\pm0.120$ & $-0.068\pm0.060$ & 12.0 & 6 & 4 & 1 \\
out2\_s29 & 159h11m56.4s & -27d33m35.3s & 33.3 & 130.7 & $4383\pm95$ & $324\pm90$ & $0.021\pm0.205$ & $-0.052\pm0.084$ & 7.7 & 6 & 4 & 1 \\
out2\_s30 & 159h11m20.4s & -27d33m23.0s & 26.6 & 127.5 & $3930\pm66$ & -- & $0.000\pm0.052$ & $0.099\pm0.044$ & 16.7 & 6 & 4 & 1 \\
out2\_s31 & 159h11m34.8s & -27d33m46.4s & 33.2 & 133.3 & $4267\pm114$ & -- & $0.007\pm0.104$ & $0.217\pm0.052$ & 10.5 & 6 & 4 & 1 \\
out2\_s32 & 159h10m51.6s & -27d33m28.1s & 26.2 & 128.8 & $3688\pm77$ & $635\pm78$ & $0.097\pm0.052$ & $0.017\pm0.088$ & 16.7 & 6 & 4 & 1 \\
out2\_s33 & 159h11m06.0s & -27d33m52.2s & 32.6 & 134.6 & $3778\pm78$ & -- & $0.062\pm0.049$ & $0.144\pm0.043$ & 15.0 & 6 & 4 & 1 \\
out2\_s34 & 159h10m04.8s & -27d33m20.5s & 25.9 & 126.8 & $4305\pm79$ & $501\pm58$ & $-0.040\pm0.109$ & $-0.121\pm0.058$ & 22.3 & 8 & 2 & 1 \\
out2\_s35 & 159h09m39.6s & -27d33m13.7s & 27.2 & 124.8 & $3931\pm57$ & $445\pm55$ & $0.113\pm0.051$ & $-0.025\pm0.101$ & 20.9 & 6 & 4 & 1 \\
out2\_s36 & 159h10m15.6s & -27d33m52.2s & 32.7 & 134.6 & $4985\pm55$ & -- & $0.254\pm0.046$ & $0.255\pm0.052$ & 27.5 & 6 & 4 & 1 \\
out2\_s37 & 159h09m21.6s & -27d33m26.6s & 32.4 & 128.4 & $4104\pm87$ & $395\pm67$ & $-0.012\pm0.155$ & $-0.091\pm0.062$ & 16.4 & 6 & 4 & 1 \\
\end{longtable}
\tablefoot{(1) Slit ID; (2) Centre of slit right ascension; (3) Centre of
slit declination; (4) Radial distance to NGC\,3311 at RA$=$10h36m42.71s
and DEC$=-$27d31m41.13s in kpc; (5) Position angle to the centre of
NGC\,3311 (from North to East); (6) Heliocentric line-of-sight recession
velocity; (7) Line-of-sight velocity dispersion; (8) Gauss-Hermite $h_3$
coefficient; (9) Gauss-Hermite $h_4$ coefficient; (10) Signal-to-Noise
per \AA; (11) Additive polynomial for continuum fit in pPXF; (12)
Multiplicative polynomial for the continuum fit in pPXF; (13) Flag:
1 - belongs to NGC\,3311, 2 - belongs to NGC\,3309, 3 - belongs to
HCC\,007.}
\end{landscape}
\end{longtab}

\onecolumn
\section{Additional figures}

\begin{figure*}[b!]
\centering
\includegraphics[width=0.90\linewidth]{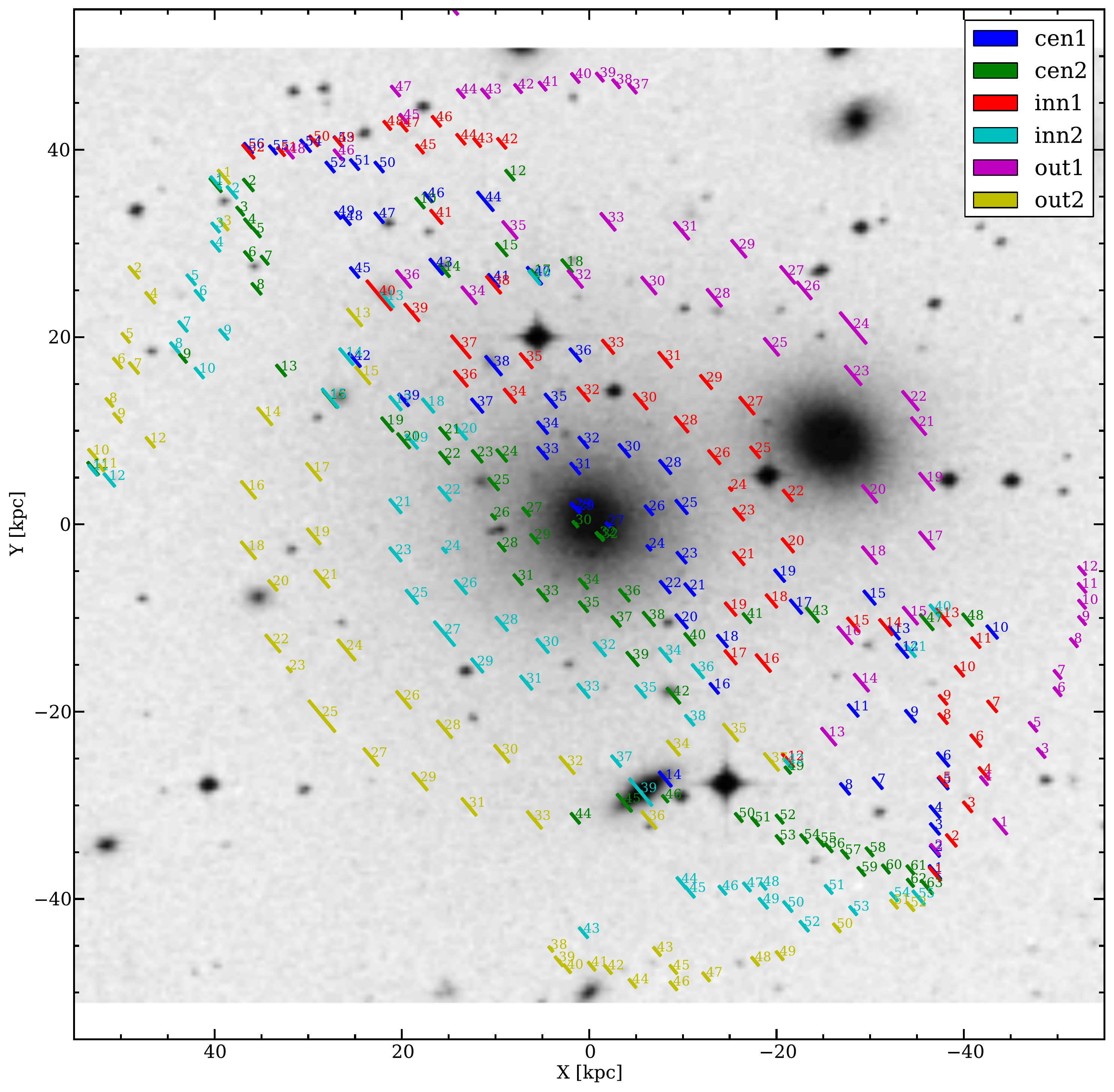}
\caption{Finding chart for slits. Different masks are colour-coded.
The names of the masks are given in the legend on the upper right.}
\label{fig:findchart}
\end{figure*}

The first additional Figure (Fig.\,\ref{fig:findchart}) highlights the
adopted onion shell observing strategy with six different mask
sets of half-rings (three on each side of the galaxy). The slit
numbers are given, and the mask names colour-coded, to ease
the identification of spectra in the text and Table\,\ref{tab:kinematics}.

\begin{figure}[ht]
\centering
\includegraphics[width=0.98\linewidth]{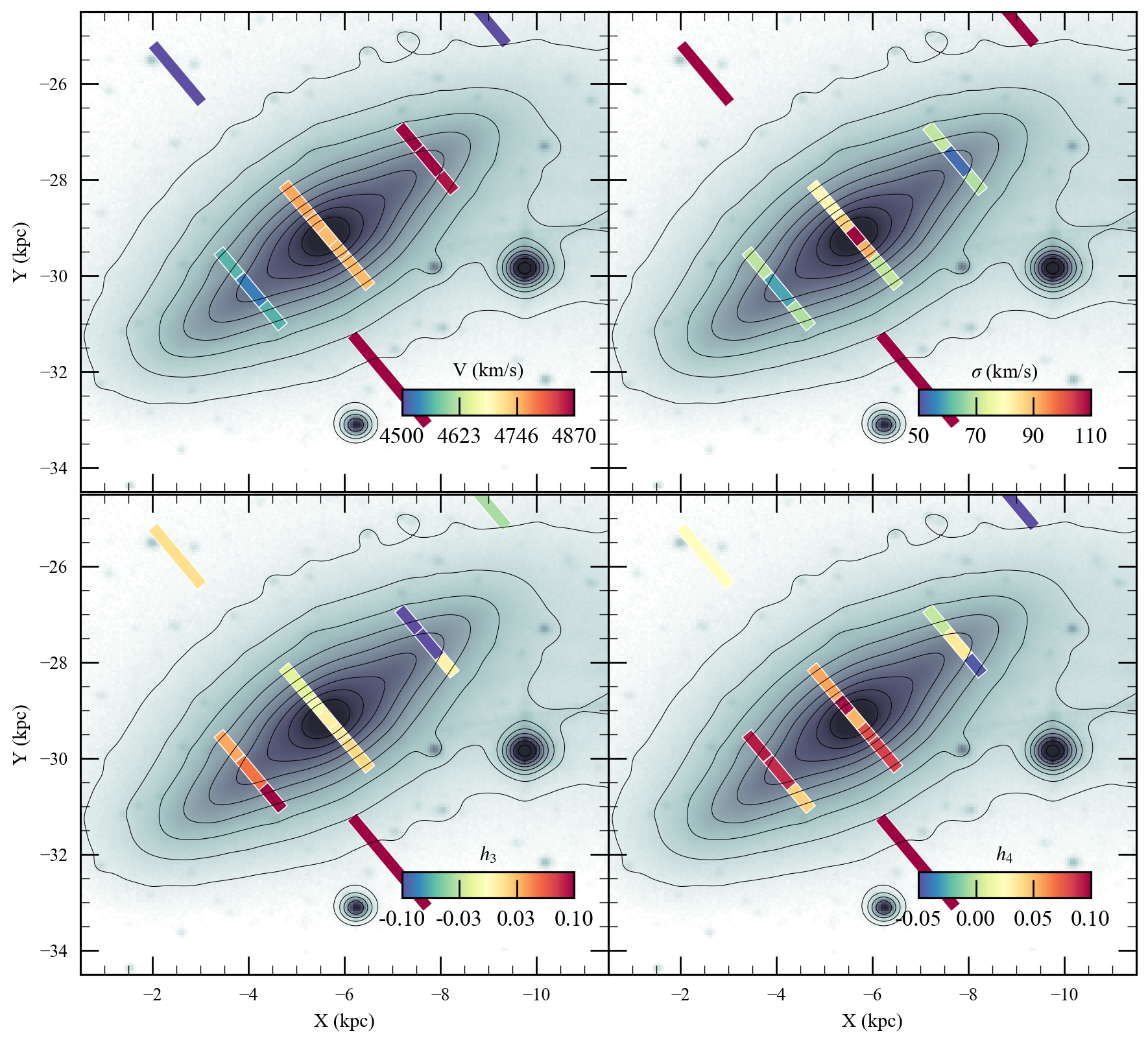}
\caption{From top left to bottom right: Line-of-sight velocity, velocity
dispersion, moments $h_3$ (skewness) and $h_4$ (kurtosis) in the
sub-sections of the  three slits on top of the lenticular galaxy HCC\,007.
Adjacent slits on NGC\,3311's halo are also shown. The rotation
signal of HCC\,007 is clearly visible in all maps.}
\label{fig:hcc007_maps}
\end{figure}

\begin{figure}[ht]
\centering
\includegraphics[width=0.98\linewidth]{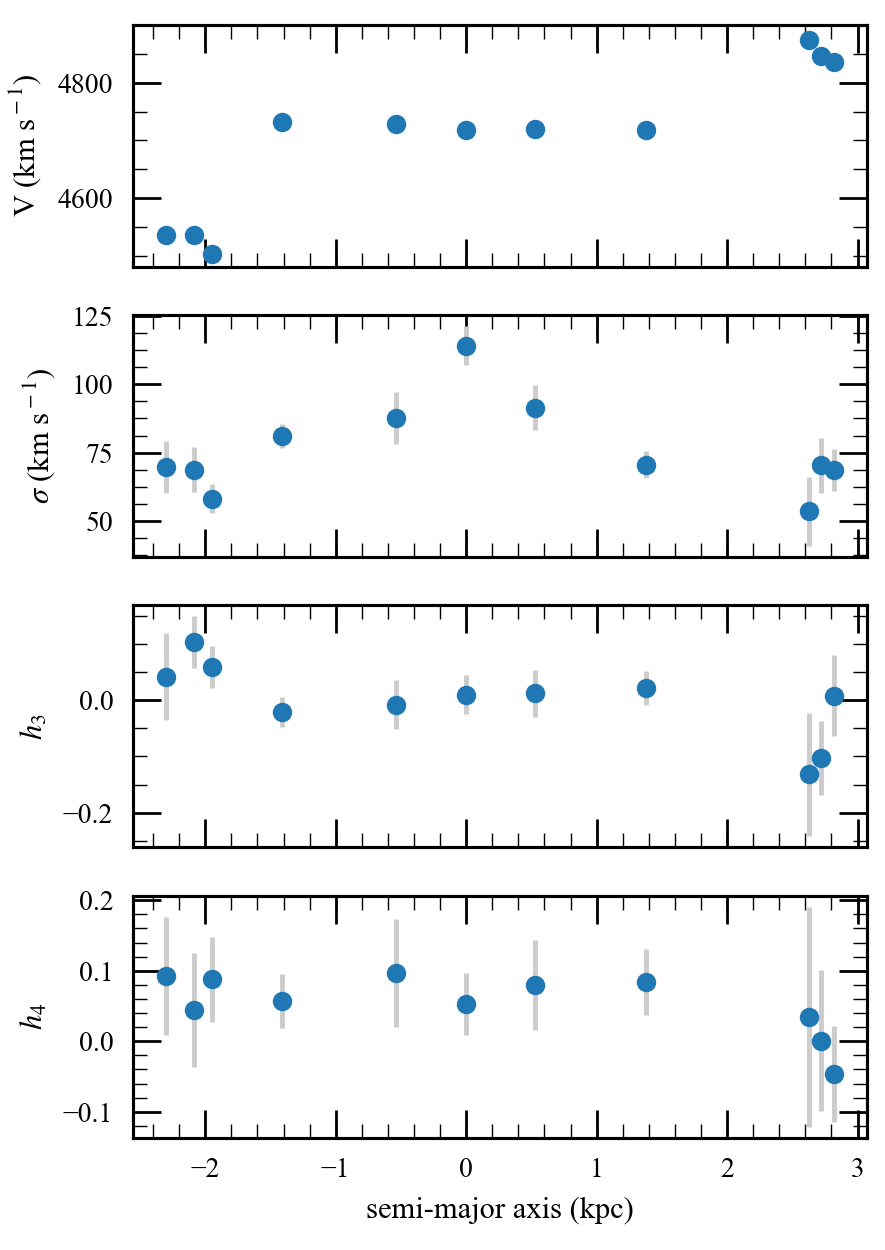}
\caption{From top to bottom: Line-of-sight velocity, velocity
dispersion, moments $h_3$ (skewness) and $h_4$ (kurtosis) as
function of isophotal distance from HCC\,007.}
\label{fig:hcc007_kinematic}
\end{figure}

Figures \ref{fig:hcc007_maps} and \ref{fig:hcc007_kinematic}
show the velocity field and LOSVD higher moments of and around
the Hydra\,I cluster lenticular galaxy HCC\,007, which seems to
interact with NGC\,3311 as judged from a wide tidal tail feature
east of the galaxy.

\twocolumn

\end{appendix}

\end{document}